\documentclass[11pt,preprint]{aastex}
\usepackage{graphicx}

\begin{document}

\title{ALMA Imaging of Millimeter/Submillimeter Continuum Emission in Orion~KL}

\author{Tomoya HIROTA\altaffilmark{1,2}, 
Mi Kyoung KIM\altaffilmark{3}, 
Yasutaka KURONO\altaffilmark{4,5},
Mareki HONMA\altaffilmark{1,2}} 
\email{tomoya.hirota@nao.ac.jp}
\altaffiltext{1}{National Astronomical Observatory of Japan, Osawa 2-21-1, Mitaka, Tokyo 181-8588, Japan}
\altaffiltext{2}{Department of Astronomical Sciences, The Graduate University for Advanced Studies (SOKENDAI), Osawa 2-21-1, Mitaka, Tokyo 181-8588, Japan}
\altaffiltext{3}{Korea Astronomy and Space Science Institute, Hwaam-dong 61-1, Yuseong-gu, Daejeon, 305-348, Republic of Korea}
\altaffiltext{4}{Chile Observatory, National Astronomical Observatory of Japan, Osawa 2-21-1, Mitaka, Tokyo 181-8588, Japan}
\altaffiltext{5}{Joint ALMA Observatory, Alonso de Cordova 3107 Vitacura, Santiago 763-0355, Chile}

\begin{abstract}
We have carried out high resolution observations with Atacama Large Millimeter/Submillimeter Array (ALMA) of continuum emission from Orion~KL region. 
We identify 11 compact sources at ALMA band~6 (245~GHz) and band~7 (339~GHz), including Hot Core, Compact Ridge, SMA1, IRc4, IRc7, and a radio source I (Source~I). 
Spectral energy distribution (SED) of each source is determined by using previous 3~mm continuum emission data. 
Physical properties such as size, mass, hydrogen number density and column density are discussed based on the dust graybody SED. 
Among 11 identified sources, Source~I, a massive protostar candidate, is a dominant energy source in Orion~KL. 
We extensively investigate its SED from centimeter to submillimeter wavelengths. 
The SED of Source~I can be fitted with a single power-law index of 1.97 suggesting an optically thick emission. 
We employ the H$^{-}$ free-free emission as an opacity source of this optically thick emission.  
The temperature, density, and mass of the circumstellar disk associated with Source~I are constrained by the SED of H$^{-}$ free-free emission. 
Still the fitting result shows a significant deviation from the observed flux densities. 
Combined with the thermal dust graybody SED to explain excess emission at higher frequency, a smaller power-law index of 1.60 for the H$^{-}$ free-free emission is obtained in the SED fitting. 
The power-law index smaller than 2 would suggest a compact source size or a clumpy structure unresolved with the present study. 
Future higher resolution observations with ALMA are essential to reveal more detailed spatial structure and physical properties of Source~I. 
\end{abstract}


\section{Introduction}

Located at a distance of 420~pc from the Sun \citep{hirota2007,menten2007,kim2008}, the Orion Kleinmann-Low \citep[KL;][]{kleinmann1967} region is known as the nearest site of active massive-star formation. 
It has been recognized as one of the best laboratories to study massive star-formation processes \citep{genzel1989,bally2008}. 
Because of its extremely high opacity, observations have been made mainly in the centimeter, millimeter/submillimeter, and infrared wavelengths. 
In particular, radio interferometers have been powerful tools to study in detail about physical properties of young stellar objects (YSOs) in Orion~KL \citep[e.g. ][]{beuther2004, beuther2005, beuther2006, tang2010,  favre2011, friedel2011, zapata2011, plambeck2013} at the highest spatial resolution. 
These observational studies have identified some of the remarkable sources, such as Becklin-Neugebauer \citep[BN;][]{becklin1967} object, Hot Core, Compact Ridge, a radio source labeled I (Source~I), infrared source labeled n (Source~n), and submillimeter source identified by Submillimeter Array (SMA), SMA1, although their properties are still a matter of debate. 

Among these compact sources, the most prominent energy source in Orion~KL is thought to be Source~I \citep{menten1995}. 
It is a candidate of massive protostar driving a so-called low velocity bipolar outflow along the northeast-southwest direction with a scale of 1000~AU traced by the thermal SiO lines and H$_{2}$O masers \citep{genzel1989,gaume1998,hirota2007,plambeck2009,zapata2012,niederhofer2012,greenhill2013,kim2014}. 
At the center of this outflow, there is a cluster of vibrationally excited SiO masers tracing a disk wind emanating from the surface of the circumstellar disk with a diameter of $\sim$100~AU \citep{kim2008,matthews2010,greenhill2013}. 
A compact radio continuum source is associated at the center of the SiO masers which is interpreted as an edge-on circumstellar disk \citep{reid2007,goddi2011,plambeck2013}. 
Such a disk-jet system is recently studied in the submillimeter H$_{2}$O lines by using a newly constructed Atacama Large Millimeter/Submillimeter Array (ALMA) \citep{hirota2012,hirota2014a}. 

Nevertheless, the nature of Source~I and associated disk/outflow system are still far from a complete understanding. 
One of the long-standing issues is an origin of the radio continuum emission associated with Source~I. 
A spectral energy distribution (SED) of Source~I from centimeter to submillimeter wavelength can be fitted to a power-law function, $F_{\nu} \propto \nu^{2}$, indicative of optically thick emission up to submillimeter wavelength \citep[][and references therein]{plambeck2013}. 
Based on such a SED of Source~I, it has been interpreted as either a combination of free-free emission and thermal dust emission \citep{beuther2004, beuther2006, reid2007, plambeck2013} or a H$^{-}$ free-free emission from a neutral atomic/molecular gas \citep{reid2007, plambeck2013}. 
Although physical properties of Source I such as mass, temperature, and ionization degree are still under
debate, more recent results support a latter scenario \citep[e.g.][]{plambeck2013}. 

In order to reveal physical properties of continuum sources in Orion~KL, in particular Source~I, we present observational study of multi-band continuum emission with the sub-arcsecond resolution by using ALMA. 
The ALMA observations open new wavelength windows with the higher spatial resolution. 
Combined with previous high-resolution observations at lower frequency bands, we will discuss about basic physical properties of some of the key sources in Orion~KL. 

\section{Observations and Data Analysis}

Observations of the millimeter and submillimeter continuum emission were carried out with ALMA in several sessions during the early science operation in the cycle~0 period  at bands 6 and 7 
(ADS/JAO.ALMA\#2011.0.00199.S). 
We also employed the ALMA Science Verification (SV) data (ADS/JAO.ALMA\#2011.0.00009.SV) at band~6. 
Details of each session is summarized in Table \ref{tab-obs}. 

\subsection{Band 6 data}

The observation in cycle~0 was done in the extended configuration on 2012 April 08 with 17$\times$12~m antennas. 
The primary beam size of each 12~m antenna is 25\arcsec. 
The baseline lengths ranged from 17 to 310~k$\lambda$ (from 21 to 385~m). 
The observed frequency ranges were 240$-$244~GHz and 256$-$260~GHz consisted of 4 spectral windows with bandwidth of 2~GHz for each. 
Dual polarization data were obtained simultaneously. 
The ALMA correlator was set for low resolution wideband continuum observations and the spectral resolution was 15.625~MHz. 
The target source was Orion~KL and the tracking center position was taken to be the bursting 22~GHz H$_{2}$O maser, RA(J2000)=05h35m14.1250s, Decl(J2000)=-05d22\arcmin36.486\arcsec \citep{hirota2011, hirota2014b}. 
The total on-source integration time was 30~s. 
A primary flux calibrator, band-pass calibrator, and secondary gain calibrator were Callisto, J053851-440507, and J0607-085, respectively. 
The system noise temperatures ranged from 102~K to 122~K depending on the spectral window. 

To improve the image quality, we combined the ALMA SV data at band~6 with our cycle~0 data. 
The SV data were obtained in the 20 frequency settings to cover full spectral emissions from 214~GHz to 247~GHz. 
We here only employed part of the frequency range from 230 to 232~GHz to achieve an image sensitivity comparable to the cycle~0 data. 
The SV data were taken on 2012 January 20 in the compact configuration with 16$\times$12~m antennas separated from 14 to 203~k$\lambda$ (from 17 to 265~m). 
The tracking center position of Orion~KL was set to be R. A. =05h35m14.35s and decl.=$-$05$^{\circ}$22\arcmin35\arcsec.0 (J2000), which is 4\arcsec \ northeast of that of the cycle 0 data. 

In this paper, we define the band~6 frequency to be 245~GHz as it is the central frequency of all the combined data. 

\subsection{Band 7 data}

For band~7, observations in two different frequency settings were carried out; continuum mode and spectral line mode. 
The tracking center position was the same as band~6 cycle~0 data; RA(J2000)=05h35m14.1250s, Decl(J2000)=-05d22\arcmin36.486\arcsec. 

Observations in the continuum mode was done in the extended configuration on 2012 October 23 with 23$\times$12~m antennas. 
The primary beam size of each 12~m antenna is 17\arcsec. 
The baseline lengths ranged from 19 to 425~k$\lambda$ (from 17 to 372~m). 
The observed frequency ranges were 341.5$-$345.4~GHz and 353.5$-$357.5~GHz consisted of four 2~GHz-bandwidth spectral windows with dual-polarization. 
The ALMA correlator was set for low resolution wideband continuum observations and the spectral resolution was 15.625~MHz. 
The total on-source integration time was 100~s. 
A primary flux calibrator, band-pass calibrator, and secondary gain calibrator were Callisto, J0423-013, and J0607-085, respectively. 
The system noise temperatures ranged from 116~K to 162~K depending on the spectral window. 

We carried out monitoring observations of the submillimeter H$_{2}$O lines at band~7 to study the H$_{2}$O maser burst event at 22~GHz in multi-frequency \citep{hirota2014a}. 
For this purpose, three sessions of spectral line observations were carried out on 2012 July 16, August 25, and October 21 during the cycle~0 period. 
Number of 12~m antennas were different from epoch to epoch; 21, 28, and 22 at the first, second, and third epochs, respectively. 
Array configurations were in the extended mode with the baseline lengths of 16-427~k$\lambda$ (15-398~m), 22-416~k$\lambda$ (21-388~m), and 18-389~k$\lambda$ (17-363~m) at the first, second, and third epoch, respectively. 
Four spectral windows were set at the frequency of 321.0-321.5~GHz, 322.1-322.6~GHz, 334.4-334.9~GHz, and 336.0-336.5~GHz with the bandwidth of 469~MHz for each. 
Spectral resolution was set to be 0.122~MHz. 
The on-source integration time of the target source was about 100~s for each session. 
A primary flux calibrator, band-pass calibrator, and secondary gain calibrator were Callisto, J053851-440507/J0423-013, and J0607-085, respectively. 
The system noise temperature ranged from 125-341~K, at which frequency around 322~GHz was significantly affected by the strong atmospheric absorption of the 325~GHz H$_{2}$O line. 

Similar to the band~6 data, we define the band~7 frequency to be 339~GHz as the central frequency of all the combined data. 

\begin{figure*}
\begin{center}
\includegraphics[width=15cm]{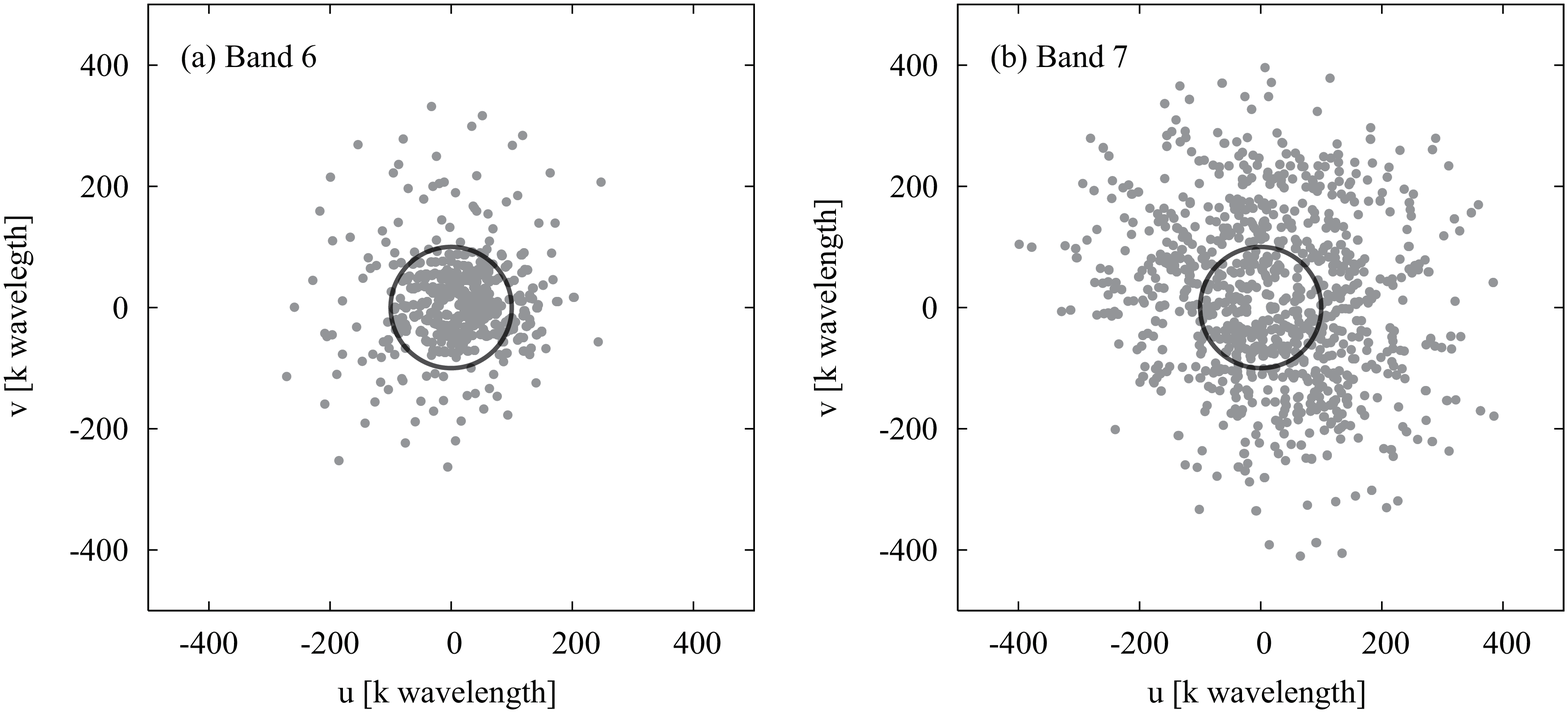}
\caption{
Full uv coverage of ALMA observations. 
(a) Band 6 data including both compact (SV) and extended (cycle 0) configurations. 
(b) Band 7 data including all of the three epochs in spectral line mode and one in continuum mode. 
The circles indicate the uv length of 100~k$\lambda$. 
}
\label{fig-uv}
\end{center}
\end{figure*}

\begin{deluxetable}{clccrcrcrr}
\rotate
\tablewidth{0pt}
\tabletypesize{\scriptsize}
\tablecaption{Summary of Observations
 \label{tab-obs}}
\tablehead{
\colhead{}                          & 
\colhead{}                          & \colhead{Center}   & 
\colhead{Total}                     & \colhead{Effective } & 
\colhead{Number}                          & 
\colhead{On-source} & 
\colhead{}                          & \colhead{}    & 
\colhead{}       
\\
\colhead{}                          & 
\colhead{}                      & \colhead{frequencies\tablenotemark{a}}   & 
\colhead{bandwidth}                 & \colhead{bandwidth} & 
\colhead{of}                 & 
\colhead{Time} & 
\colhead{FWHM}                      & \colhead{PA}    & 
\colhead{rms} 
\\
\colhead{Band}                      & 
\colhead{Date (in 2012)}                 & \colhead{[GHz]}   & 
\colhead{[MHz]}                     & \colhead{[MHz]} & 
\colhead{Antennas}                  & 
\colhead{[sec]} & 
\colhead{[arcsec]}                  & \colhead{[degree]}    & 
\colhead{[mJy~beam$^{-1}$]}
}
\startdata
6
  & Jan. 20                     & 231     & 1875    &  167    &      16 &  1196.1  & 1.76$\times$1.19 &  \ -1 & 19 \\
  & Apr. 08                     & 250 & 8000    & 3547    &      17 &    30.3  & 0.68$\times$0.50 &   -75 & 5 \\
  & All uv                        & 245 & \nodata & \nodata & \nodata &  \nodata & 0.81$\times$0.64 &   -70 & 9 \\
  & uv$>$100~k$\lambda$      & 245 & \nodata & \nodata & \nodata &  \nodata & 0.63$\times$0.47 &   -72 & 5 \\
7
  & Oct. 23                     & 350 & 8000    & 3641    &      23 &    100.8  & 0.47$\times$0.44 & \ \ 0 & 9 \\
  & Jul. 16, Aug. 25, Oct. 21\tablenotemark{b}
                                    & 329 & 1875    &  648    & 21,28,22 &  326.8  & 0.55$\times$0.46 &  \ 59 & 12 \\
  & All uv                        & 339 & \nodata & \nodata & \nodata &  \nodata & 0.49$\times$0.45 &  \ 59 & 9 \\
  & uv$>$100~k$\lambda$      & 339 & \nodata & \nodata & \nodata &  \nodata & 0.46$\times$0.41 &  \ 57 & 5 \\
\enddata
\tablenotetext{a}{Center frequencies for LSB (lower side band) and USB (upper side band). \\
Only one spectral window in LSB is employed for band~6 data on January 20, 2012 (SV data). }
\tablenotetext{b}{Combine three epochs for spectral line monitoring at band~7. }
\end{deluxetable}

\subsection{Synthesis imaging}

The data were calibrated and imaged with the Common Astronomy Software Applications (CASA) package. 
For the band~6 data, we combined calibrated SV and cycle~0 data by using the task {\tt{concat}} in CASA. 
Plot of uv coverages of full-array are shown in Figure \ref{fig-uv}. 
For band~7 data, we combined all of the calibrated data for both in continuum and spectral line mode by using the CASA task {\tt{concat}}. 
After combining the data, continuum images of Orion~KL were made using the CASA task {\tt{clean}}. 
The line emissions were excluded to make continuum images by integrating over the line-free channels. 
Resultant effective bandwidths were almost half of the observed frequency range as listed in Table \ref{tab-obs}. 
Both phase and amplitude self-calibrations were done with the continuum images by the CASA task {\tt{gaincal}} and these results were applied to the visibility data by the CASA task {\tt{applycal}}. 
The resultant image rms noise level of the continuum emission and the uniform-weighted synthesized beam size were summarized in Table \ref{tab-obs}. 

\begin{figure*}
\begin{center}
\includegraphics[width=15cm]{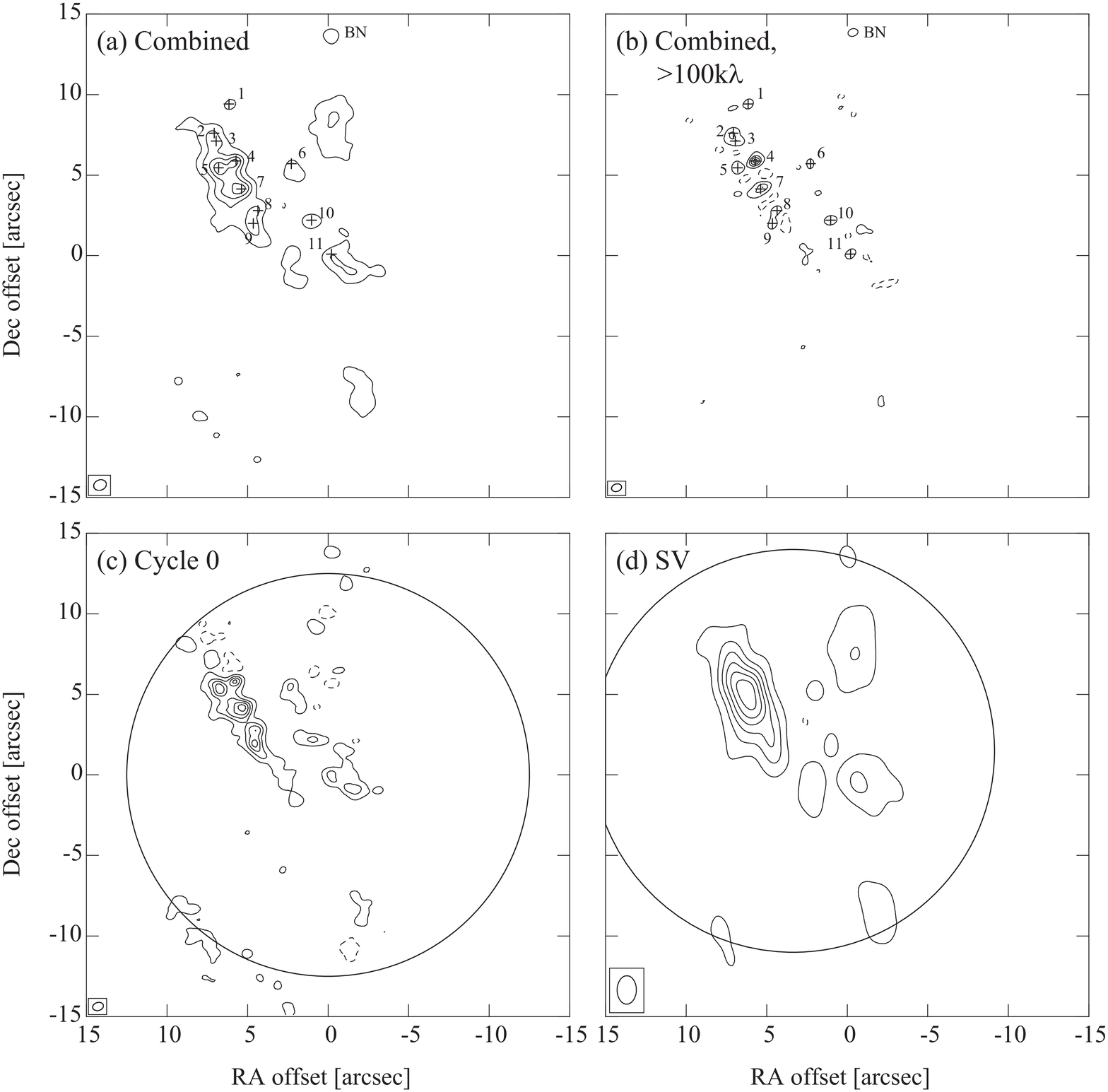}
\caption{
Continuum emission maps for band~6 data. 
Synthesized beam size is indicated at the bottom-left corner in each panel. 
The contours start at -5$\sigma$ level with an interval of 10$\sigma$ (5, 15, 25, ...). 
The (0,0) position is the tracking center position of the cycle~0 data, RA(J2000)=05h35m14.1250s, Decl(J2000)=-05d22\arcmin36.486\arcsec. 
The HKKH source ID numbers are indicated in panels (a) and (b). 
(a) Combine both the SV and cycle~0 data. 
The noise level (1$\sigma$) is 9~mJy~beam$^{-1}$. 
(b) Same as (a) but excluding those with uv length less than 100~k$\lambda$. 
The noise level (1$\sigma$) is 5~mJy~beam$^{-1}$. 
(c) Cycle 0 data. 
The noise level (1$\sigma$) is 5~mJy~beam$^{-1}$. 
A circle indicates a primary beam size of the 12~m antenna, 25\arcsec. 
(d) SV data. 
The noise level (1$\sigma$) is 19~mJy~beam$^{-1}$. 
A circle indicates a primary beam size of the 12~m antenna, 25\arcsec.
Note that the pointing center is 4\arcsec \ northeast of that of Cycle 0 data (see panel (c)). 
}
\label{fig-mapb6}
\end{center}
\end{figure*}

\begin{deluxetable}{lllccrr}
\tablewidth{0pt}
\tabletypesize{\scriptsize}
\tablecaption{Continuum sources at band~6
\label{tab-band6}}
\tablehead{
\colhead{ID}                 & 
\colhead{$\alpha$(J2000)}  & \colhead{$\delta$(J2000)}  &
\colhead{Convolved size}   & \colhead{PA}   & 
\colhead{Integrated}       & \colhead{Peak} \\
\colhead{HKKH}                              &
\colhead{+05h35m (s)}                     & \colhead{-05d22\arcmin (\arcsec)}         & 
\colhead{(\arcsec $\times$ \arcsec)}      & \colhead{(deg)}                           &
\colhead{(mJy)}                           & \colhead{(mJy~beam$^{-1}$)}
}
\startdata
 1  & 14.5373(13) &  27.072(15) & 0.82(3) $\times$ 0.61(4)  & 127(3)  &  123(8)  &  74(5)   \\
 2  & 14.5988(26) &  28.876(39) & 0.73(6) $\times$ 0.48(8)  & 110(4)  &  123(16) & 106(13)  \\
 3  & 14.5902(27) &  29.369(31) & 0.88(6) $\times$ 0.52(8)  &  97(50) &  160(21) & 104(13)  \\
 4  & 14.5073(10) &  30.603(12) & 0.66(2) $\times$ 0.47(3)  & 124(2)  &  284(14) & 272(13)  \\
 5  & 14.5804(26) &  31.029(31) & 0.61(6) $\times$ 0.57(8)  & 97(31)  &  118(15) & 102(13)  \\
 6  & 14.2775(20) &  30.776(23) & 0.68(6) $\times$ 0.46(5)  & 170(4)  &   53(5)  &  51(5)   \\
 7  & 14.4853(20) &  32.338(23) & 1.09(5) $\times$ 0.51(6)  & 121(3)  &  267(25) & 141(13)  \\
 8  & 14.4162(66) &  33.684(77) & 0.32(17)$\times$ 0.20(19) &  74(29) &   13(4)  &  58(18)  \\
 9  & 14.4360(78) &  34.488(91) & 0.47(18)$\times$ 0.35(24) &  99(19) &   27(10) &  49(18)  \\
10 & 14.1939(16) &  34.289(19) & 0.65(4) $\times$ 0.44(5)  & 100(4)  &   69(5)  &  71(6)   \\
11 & 14.1112(18) &  36.383(21) & 0.66(4) $\times$ 0.48(5)  & 134(5)  &   57(5)  &  52(4)   \\
\enddata
\tablecomments{Numbers in parenthesis represent fitting errors determined by the CASA task {\tt{imfit}} in unit of the last significant digits. }
\end{deluxetable}

\begin{figure*}
\begin{center}
\includegraphics[width=15cm]{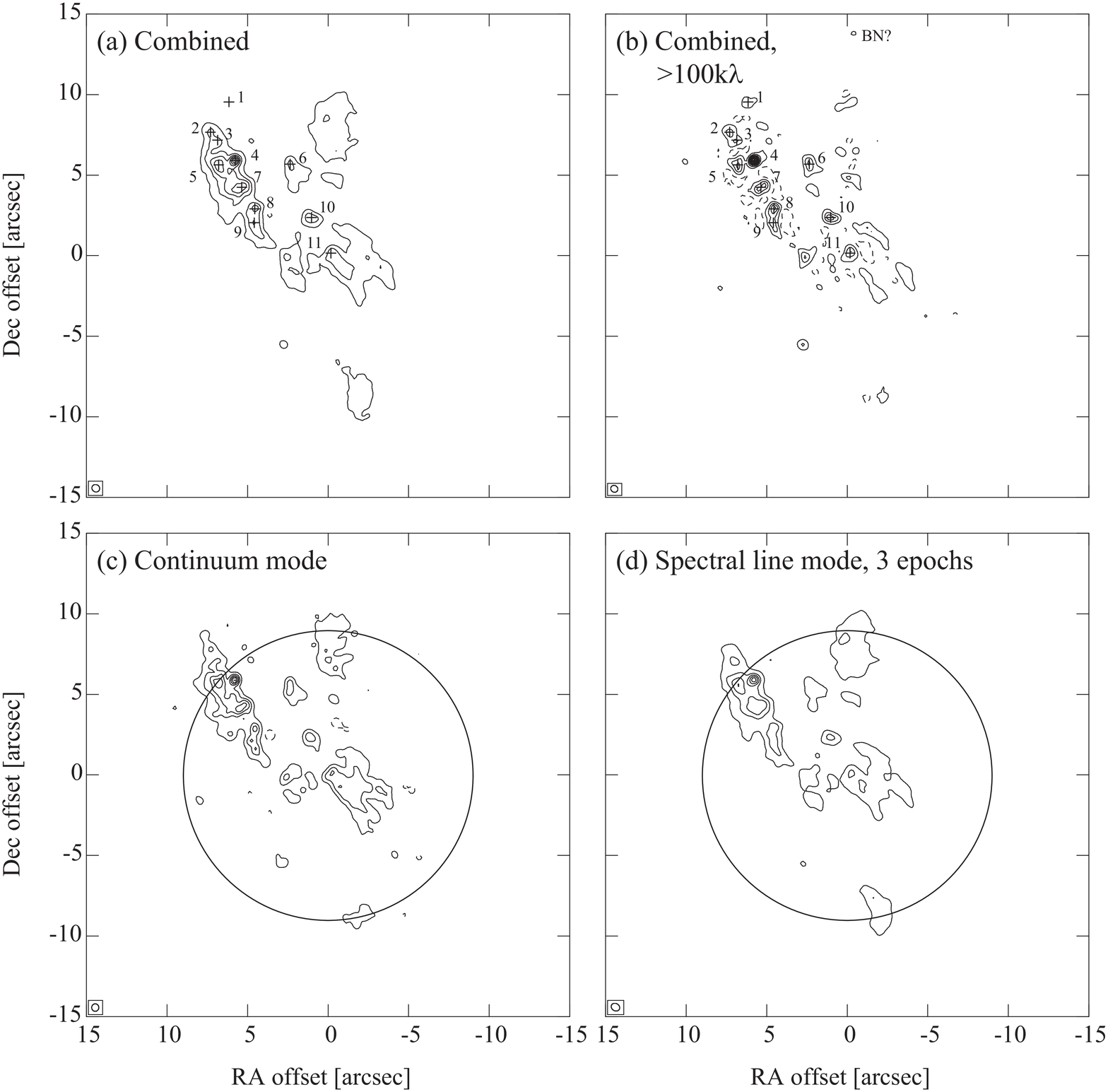}
\caption{
Continuum emission maps for band~7 data. 
Synthesized beam size is indicated at the bottom-left corner in each panel. 
The contours start at -5$\sigma$ level with an interval of 10$\sigma$ (5, 15, 25, ...). 
The (0,0) position is the tracking center position of the cycle~0 data, RA(J2000)=05h35m14.1250s, Decl(J2000)=-05d22\arcmin36.486\arcsec. 
The HKKH source ID numbers are indicated in panels (a) and (b). 
(a) Combine all of the four epochs including both in spectral line mode and continuum mode. 
The noise level (1$\sigma$) is 9~mJy~beam$^{-1}$. 
(b) Same as (a) but excluding those with uv length less than 100~k$\lambda$. 
The noise level (1$\sigma$) is 5~mJy~beam$^{-1}$. 
(c) A single epoch data observed in the continuum mode. 
The noise level (1$\sigma$) is 9~mJy~beam$^{-1}$. 
A circle indicates a primary beam size of the 12~m antenna, 18\arcsec. 
(d) Combine three epochs of observations in spectral line mode. 
The noise level (1$\sigma$) is 12~mJy~beam$^{-1}$. 
A circle indicates a primary beam size of the 12~m antenna, 18\arcsec. 
}
\label{fig-mapb7}
\end{center}
\end{figure*}

\begin{deluxetable}{lllccrr}
\tablewidth{0pt}
\tabletypesize{\scriptsize}
\tablecaption{Continuum sources at band~7
\label{tab-band7}}
\tablehead{
\colhead{ID}                 & 
\colhead{$\alpha$(J2000)}  & \colhead{$\delta$(J2000)}  &
\colhead{Convolved size}   & \colhead{PA}   & 
\colhead{Integrated}       & \colhead{Peak} \\
\colhead{HKKH}                              &
\colhead{+05h35m (s)}                     & \colhead{-05d22\arcmin (\arcsec)}         & 
\colhead{(\arcsec $\times$ \arcsec)}      & \colhead{(deg)}                           &
\colhead{(mJy)}                           & \colhead{(mJy~beam$^{-1}$)}
}
\startdata
 1  & 14.5372(17) &  26.939(24) & 0.93(5) $\times$ 0.66(5)  & 104(5)  &  505(58) & 156(18)  \\
 2  & 14.6141(17) &  28.816(24) & 0.65(5) $\times$ 0.57(5)  &  33(3)  &  591(66) & 299(33)  \\
 3  & 14.5844(23) &  29.317(42) & 0.44(10)$\times$ 0.33(7)  & 131(7)  &  161(25) & 212(33)  \\
 4  & 14.5113(5) &  30.571(7)  & 0.50(2) $\times$ 0.44(1)  & 126(5)  & 1001(34) & 849(29)  \\
 5  & 14.5797(21) &  30.852(30) & 0.69(6) $\times$ 0.58(6)  &  16(12) &  639(91) & 301(43)  \\
 6  & 14.2827(09) &  30.801(13) & 0.91(3) $\times$ 0.54(3)  &  10(2)  &  487(29) & 183(11)  \\
 7  & 14.4837(18) &  32.238(26) & 0.83(5) $\times$ 0.51(5)  & 121(4)  &  534(64) & 237(29)  \\
 8  & 14.4297(16) &  33.582(24) & 0.62(5) $\times$ 0.51(5)  & 141(8)  &  396(44) & 238(26)  \\
 9  & 14.4323(29) &  34.440(41) & 0.63(8) $\times$ 0.63(9)  &  69(10) &  281(55) & 135(26)  \\
10 & 14.1938(10) &  34.116(14) & 0.75(3) $\times$ 0.47(3)  &  72(2)  &  336(22) & 176(11)  \\
11 & 14.1131(16) &  36.327(23) & 0.70(5) $\times$ 0.56(5)  & 165(7)  &  271(29) & 129(14)  \\
\enddata
\tablecomments{Numbers in parenthesis represent fitting errors determined by the CASA task {\tt{imfit}} in unit of the last significant digits. }
\end{deluxetable}

\section{Results}

Figures \ref{fig-mapb6} and \ref{fig-mapb7} show continuum images of Orion~KL at bands~6 and 7, respectively, obtained with various uv coverages. 
To achieve higher image quality and sensitivity, we combined different datasets observed in different uv coverage as explained in previous section. 
Overall spatial structures of the band~6 and band~7 images with full uv coverage (Figures \ref{fig-mapb6}(a) and \ref{fig-mapb7}(a)) appear quite similar with each other. 
The main emission region shows an elongated ridge-like structure \citep[``main dust ridge'' in ][]{tang2010} along the northeast-southwest direction consisted of Source~I, Hot Core and SMA1 as labeled in Figure \ref{fig-label}. 
At western side of this main dust ridge and south of the BN object, there are several diffuse emission components located along the north-south direction, which can be identified as the Northwest Clump and Compact Ridge (Figure \ref{fig-label}). 
Our ALMA images trace part of these extended emission corresponding to compact highest density condensations. 

In order to identify compact continuum sources, we made synthesized images by employing only baselines longer than 100~k$\lambda$ as shown in Figures \ref{fig-mapb6}(b) and  \ref{fig-mapb7}(b). 
The extended emission components are resolved out in our ALMA images (Figures \ref{fig-mapb6}(b) and \ref{fig-mapb7}(b)) because of the lack of short baselines in the ALMA extended configuration. 
Although the high resolutions images (Figures \ref{fig-mapb6}(b) and \ref{fig-mapb7}(b)) show lower intensities than those of full uv coverages (Figures \ref{fig-mapb6}(a) and \ref{fig-mapb7}(a)), compact sources are clearly detected with sufficiently high signal-to-noise ratios greater than 5$\sigma$. 
Note that there are negative contour levels in these images due to insufficient uv coverages for shorter baselines. 
When we employ the visibility data with baselines longer than 200~k$\lambda$, only Source~I remains unresolved implying a compact structure \citep{beuther2004}. 

For band~6 data, we compare the cycle~0 data (Figure \ref{fig-mapb6}(c)) and SV data  (Figure \ref{fig-mapb6}(d)) obtained in the extended and compact configurations, respectively. 
Only the higher resolution cycle~0 data resolve the compact cores, as can be seen in the main dust ridge (Figure \ref{fig-mapb6}(c)). 
In contrast, we cannot see extended NW clump in the cycle~0 data which is evident in the SV data (Figure \ref{fig-mapb6}(d)). 
By combining both data (Figures \ref{fig-mapb6}(a)), we can significantly improve source structure by recovering both extended and compact features. 

Since we have carried out three epochs of monitoring observations at band~7 in the spectral line modes as well as a single-epoch continuum observation, we compared these four observational results. 
For the spectral line mode, we notice an apparent difference in the synthesize images between epoch to epoch. 
Because such a difference is not significant for the most compact condensation identified as Source~I, we attribute the reason to the difference in the uv coverage. 
When compared a image from the single-epoch continuum observation (Figure \ref{fig-mapb7}(c)) with that of combined image of all three epochs observed in the spectral line mode (Figure \ref{fig-mapb7}(d)), both results are in good agreement with each other.  
Taking into account these effects, we estimate the accuracy of the flux measurement to be 20\%. 

We determine positions and intensities of the compact cores by fitting the two-dimensional Gaussian on the images made with uv coverages of $>$100~k$\lambda$ (Figures \ref{fig-mapb6}(b) and \ref{fig-mapb7}(b)) by using the CASA task {\tt{imfit}}. 
The flux densities are measured by employing the images after correction of primary beam attenuation. 
Because there could be false detection due to sidelobes or noises, we conservatively regard the sources as real detections if their signal-to-noise ratio greater than 5 for both bands~6 and 7. 
As a result, we identify total 11 sources as listed in Tables \ref{tab-band6} and \ref{tab-band7}. 
In this paper, we define the source name as HKKH (an acronym of the authors' initials) followed by the ID number. 

For the band~6 images, the missing flux in the highest resolution images with uv coverage of $>$100~k$\lambda$ (Figure \ref{fig-mapb6}(b)) compared with the image with full-uv coverage (Figure \ref{fig-mapb6}(a)) are larger than 90\% for sources HKKH1, HKKH6, and HKKH9. 
In contrast, more than 90\% of flux densities at band~6 are recovered for sources HKKH2, HKKH4, and HKKH10, implying compact structures. 
For the band~7 images, a fraction of the missing flux in the highest resolution images with uv coverage of $>$100~k$\lambda$ (Figure \ref{fig-mapb7}(b)) is less than 50\% in comparison with those of full-uv coverage  (Figure \ref{fig-mapb7}(a)). 
This difference in the degree of missing flux between bands~6 and 7 would be attributed to the different uv coverages and beam sizes. 

For band~7, source HKKH1, HKKH2, and HKKH3 are outside of the primary beamsize of the ALMA 12~m antenna while they are within the field of view of band~6. 
Although flux calibration is not reliable for these sources, we regard them as detection both in bands~6 and 7. 
Similar to these sources, source HKKH5 is detected at the edge of the field of view of band~7 images. 
In addition, the BN object, a famous massive YSO in this region, is outside of the field of view of bands~6 and 7, as discussed later. 

\begin{deluxetable}{rrrrl}
\tablewidth{0pt}
\tabletypesize{\scriptsize}
\tablecaption{Identification of continuum sources
\label{tab-cont}}
\tablehead{
\colhead{ID} & \colhead{$I_{\mbox{339~GHz}}$} & \colhead{$I_{\mbox{245~GHz}}$} & \colhead{$I_{\mbox{90~GHz}}$\tablenotemark{a}} & \colhead{}  \\
\colhead{HKKH} & \colhead{(mJy~beam$^{-1}$)} & \colhead{(mJy~beam$^{-1}$)} & \colhead{(mJy~beam$^{-1}$)}  & \colhead{Note} 
}
\startdata
 1 &  156(18) &   74(5)  & 4.89(29)   &   C13                  \\
 2 &  299(33) &  106(13) & 4.39(43)   &   C19, MF10?           \\
 3 &  212(33) &  104(13) & \nodata    &   C33?, MF10?          \\
 4 &  849(29) &  272(13) & 50.09(97) & C20, MF11?, Source~I  \\
 5 &  301(43) &  102(13) & 7.08(94)   &   C18                  \\
 6 &  183(11) &   51(5) \ & 4.18(9) \   &   C23 IRc7            \\
 7 &  237(29) &  141(13) & 6.05(60)   &   C21, MF6, SMA1      \\
 8 &  238(26) &   58(18) & 5.31(75)   &   C22                  \\
 9 &  135(26) &   49(18) & \nodata    &   C34?, MF2            \\
10 &  176(11) &   71(6)  & \nodata    &   MF12, IRc4           \\
11 &  129(14) &   52(4)  & \nodata    &   C32, MF1, Source~R \\
\enddata
\tablenotetext{a}{See Table 1 in \citet{friedel2011}. }
\tablecomments{Numbers in parenthesis represent the errors in unit of the last significant digits. 
MF and C in column 5 are methyl formate \citep{favre2011} and 3~mm continuum \citep{friedel2011} peaks, respectively. }
\end{deluxetable}

\begin{deluxetable}{rcccccccccc}
\tablewidth{0pt}
\tabletypesize{\scriptsize}
\tablecaption{graybody fitting results for continuum sources
\label{tab-gray}}
\tablehead{
\colhead{} & \colhead{} & 
\colhead{} & \colhead{20~K} & \colhead{} & \colhead{} & 
\colhead{} & 
\colhead{} & \colhead{100~K} & \colhead{} & \colhead{} \\
\cline{3-6} \cline{8-11}
\colhead{ID} & \colhead{Diameter} & 
\colhead{$M$} & \colhead{} & \colhead{$n$(H$_{2}$)} & \colhead{$N$(H$_{2}$)} & 
\colhead{} & 
\colhead{$M$} & \colhead{} & \colhead{$n$(H$_{2}$)} & \colhead{$N$(H$_{2}$)} \\
\colhead{HKKH} & \colhead{(AU)} & 
\colhead{($M_{\odot}$)} & \colhead{$\beta$\tablenotemark{a}} & \colhead{(cm$^{-3}$)} & \colhead{(cm$^{-2}$)} & 
\colhead{} & 
\colhead{($M_{\odot}$)} & \colhead{$\beta$\tablenotemark{a}} & \colhead{(cm$^{-3}$)} & \colhead{(cm$^{-2}$)} 
}
\startdata
 1   & 313  &  0.077(3)   &   0.77(2)  &  8.6 10$^{8}$  &  4.0 10$^{24}$ & &   0.009(1)  &   0.59(5)     & 1.0 10$^{8}$   &  4.7 10$^{23}$   \\   
 2   & 252  &  0.274(35)  &   1.30(7)  &  5.8 10$^{9}$  &  2.2 10$^{25}$ & &   0.031(2)  &   1.12(4)     & 6.6 10$^{8}$   &  2.5 10$^{24}$   \\   
 3   & 213  &  0.082       &   0.58  &  2.9 10$^{9}$  &  9.2 10$^{24}$ & &   0.008  &   0.27     & 2.8 10$^{8}$   &  9.0 10$^{23}$   \\   
 4   & 215  &  0.156(119)  &   0.19(40)  &  5.4 10$^{9}$  &  1.7 10$^{25}$ & &   0.018(12)  &  -0.00(37)  \ & 6.2 10$^{8}$   &  2.0 10$^{24}$   \\   
 5   & 257  &  0.159(57)  &   0.92(19)  &  3.2 10$^{9}$  &  1.2 10$^{25}$ & &   0.018(5)  &   0.74(16)     & 3.6 10$^{8}$   &  1.4 10$^{24}$  \\   
 6   & 263  &  0.086(55)  &   0.90(33)  &  1.6 10$^{9}$  &  6.4 10$^{24}$ & &   0.010(6)  &   0.71(30)     & 1.9 10$^{8}$   &  7.4 10$^{23}$   \\   
 7   & 293  &  0.175(78)  &   0.99(23)  &  2.4 10$^{9}$  &  1.1 10$^{25}$ & &   0.020(10)  &   0.80(26)     & 2.7 10$^{8}$   &  1.2 10$^{24}$   \\   
 8   & 158  &  0.103(86)  &   0.89(43)  &  8.9 10$^{9}$  &  2.1 10$^{25}$ & &   0.012(9)  &   0.70(40)     & 1.0 10$^{9}$   &  2.4 10$^{24}$   \\   
 9   & 212  &  0.168  &   1.51  &  6.0 10$^{9}$  &  1.9 10$^{25}$ & &   0.016  &   1.19     & 5.7 10$^{8}$   &  1.8 10$^{24}$   \\   
10   & 237  &  0.145  &   1.16  &  3.7 10$^{9}$ &  1.3 10$^{25}$ & &   0.014  &   0.87     & 3.6 10$^{8}$   &  1.3 10$^{24}$   \\   
11   & 249  &  0.107  &   1.19  &  2.4 10$^{9}$ &  8.8 10$^{24}$ & &   0.010  &   0.87     & 2.2 10$^{8}$   &  8.2 10$^{23}$   \\   
\enddata
\tablenotetext{a}{Dust opacity index in equation (\ref{eq-graykappa}). }
\tablecomments{Numbers in parenthesis represent fitting errors (1$\sigma$) in unit of the last significant digits. 
If the 90~GHz data is not available, fitting errors cannot be estimated. }
\end{deluxetable}

\section{Discussion}

\subsection{Comparison with previous observational results}

\begin{figure*}
\begin{center}
\includegraphics[width=15cm]{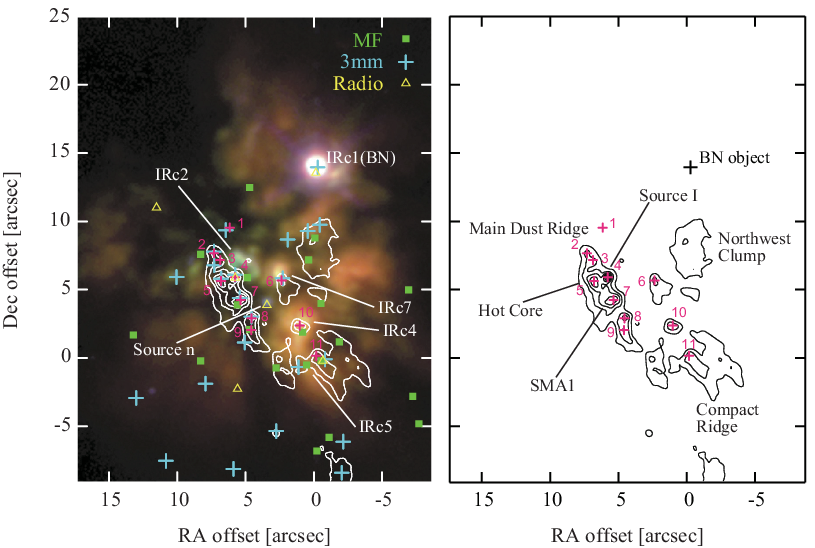}
\caption{
ALMA band~7 continuum maps (contour) superposed on the Subaru mid-infrared image \citep{okumura2011}. 
Positions of band~7 continuum sources (Table \ref{tab-band7}) are indicated by magenta smaller crosses with the HKKH ID numbers. 
Peaks of methyl formate \citep[MF;][]{favre2011}, 3mm~continuum emission \citep{friedel2011}, and radio continuum emission \citep{felli1993a,felli1993b} are indicated by green squares, blue larger crosses, and yellow triangles, respectively. 
The (0,0) position is that of the 22~GHz H$_{2}$O maser burst in the Compact Ridge \citep{hirota2011, hirota2014b}. 
}
\label{fig-label}
\end{center}
\end{figure*}

There have been number of high-resolution observations of Orion~KL with continuum emissions in wide range of wavelengths. 
We compare some of the highest resolution data reported to date. 
Examples are shown in Figure \ref{fig-label}. 
Positions of most of our detected sources are coincident with previous interferometric maps of the highest resolution millimeter continuum emission \citep{friedel2011}, molecular lines of methyl formate, HCOOCH$_{3}$ \citep{favre2011}, and radio continuum emission \citep{felli1993a,felli1993b} as shown in Figure \ref{fig-label} and Table \ref{tab-cont}. 

We compare spatial structure of our millimeter/submillimeter maps with the mid-infrared emission observed with the Subaru telescope \citep{okumura2011}. 
As can be seen in Figure \ref{fig-label}, the mid-infrared emission shows an anti-correlation with those of millimeter/submillimeter. 
It is easily interpreted that the millimeter/submillimeter emission are emitted predominantly from dust cloud with high opacity even at mid-infrared bands. 

We could not detect some of possible counterparts of the millimeter/submillimeter continuum sources detected in other tracers (Figure \ref{fig-label}). 
There are possible reasons for such differences. 
First, we employ a rather conservative detection criterion that the continuum sources are regarded as detection if their peak intensities greater than 5 times the noise level. 
This is because we eliminate contributions from strong sidelobes as possible. 
As a result, we might miss weak emission in our map. 
Furthermore, we employ the uv lengths longer than 100~k$\lambda$ to search for only compact sources. 
Thus, our continuum images significantly resolves out weak extended emission sources detected in the shorter-baselines in previous interferometer observations. 

\subsection{Graybody fitting of spectral energy distributions (SED)}

As tabulated in Table \ref{tab-cont}, we identify 11 continuum sources detected both in bands 6 and 7 images with the uv coverage of greater than 100~k$\lambda$. 
It is known that the SED of millimeter/submillimeter continuum emission associated with YSOs can be well fitted by the optically thin graybody function as a result of thermal dust emission. 
Thus we fit the observed flux densities at 245~GHz, and 339~GHz as listed in Table \ref{tab-cont} to the graybody function; 
\begin{eqnarray}
F_{\nu}^{\rm{gb}} & = & \frac{M_{\rm{tot}}}{D^{2}} \kappa_{\nu} B_{\nu}(T_{\rm{dust}}) \label{eq-gray} \\
\kappa_{\nu} & = & \kappa_{\nu_{0}} \left( \frac{\nu}{\nu_{0}} \right) ^{\beta} \label{eq-graykappa}
\end{eqnarray}
where $M_{\rm{tot}}$ is the total mass, $D$ the distance, $\kappa_{\nu}$ the total mass opacity, and $B_{\nu}(T_{\rm{dust}})$ the Planck function at the dust temperature of $T_{\rm{dust}}$. 
The total mass opacity $\kappa$ is expressed as a power law function of the frequency, and $\kappa_{\nu_{0}}=0.1$~cm$^{2}$~g$^{-1}$ at $\nu_{0}=1.2$~THz, or wavelength of 250$\mu$m, is adopted \citep{hildebrand1983} assuming the gas to dust mass ratio of 100. 
If high resolution millimeter continuum data at 90~GHz are available \citep{friedel2011}, we also employ these results in the SED fitting. 

The results of the graybody fitting are summarized in Table \ref{tab-gray}. 
The power law index of the dust opacity, $\beta$ is fitted by using two or three data points as listed in Table \ref{tab-cont}. 
Because the dust temperatures cannot be determined from our continuum observations, we assume the temperature of 100~K as obtained from the molecular line observations \citep{favre2011} and lower value of 20~K \citep{eisner2008} for comparison. 
The derived mass is anti-correlated with the temperature as a function of $(\exp (h\nu/kT) - 1)$, and hence, the lower temperature values would yield an upper limit of the mass. 
The dust opacity index, $\beta$ is close to 1 for most of the sources, except for Source~I. 
The SED of Source~I cannot be fitted by a simple optically thin thermal dust emission as discussed later. 

The origin of compact continuum sources are attributed to either star-forming or starless cores, although detailed nature of each source is still unknown (see discussion about individual sources). 
If the source is associated with embedded YSO, the compact structure of 200-300~AU and high density of 10$^{8}$-10$^{9}$~cm$^{-3}$ are indicative of circumstellar disks as an one of the possible origins of millimeter/submillimeter sources. 
The derived masses (0.08$M_{\odot}$-0.27$M_{\odot}$ for assumed temperature of 20~K) correspond to the smaller end of the masses for disk candidates in another massive star-forming region NGC6334I(N) \citep{hunter2014}. 
Due to the closer distance of Orion~KL than that of NGC6334I(N), 1.3~kpc, derived source sizes are significantly smaller than those in NGC6334I(N) \citep{hunter2014}. 
On the other hand, the derived mass range is smaller than the typical circumstellar disk masses associated with massive YSOs \citep{cesaroni2007} but is larger than those of low-mass YSOs including members of Orion Nebula Cluster \citep{eisner2008}. 
The smaller circumstellar mass compared with another massive YSO was also recognized for Source~I in Orion~KL by the previous SMA observation \citep{beuther2004}. 
If our millimeter/submillimeter sources trace the circumstellar disk of massive YSOs, the lower-mass range could imply a relatively later evolutionary stages in massive disk formation as proposed in \citet{beuther2004}. 
Nevertheless, we cannot rule out possibilities of an envelope around YSO, shocked dense gas, or starless dense gas, which are significantly resolved with our high-resolution images, or a disk associated with low-/intermediate-mass YSO. 
Key parameters to distinguish these possibilities are gas temperature and velocity structure traced by molecular lines in multiple transitions to investigate physical and dynamical properties of the continuum sources. 

\section{Individual Sources}

\subsection{Source~I (HKKH4)}
\label{sec-sourceI}

Source~I is thought to be a dominant energy source in Orion~KL \citep{menten1995}. 
Number of observational studies have been done for Source~I to reveal its nature whereas it can be detected only at longer wavelength than submillimeter band due to an extremely high opacity \citep{greenhill2004, okumura2011, sitarski2013}. 
It can be identified as a color temperature peak derived from mid-infrared images \citep{okumura2011}. 
Position of Source~I corresponds to the 3~mm continuum source C20 \citep{friedel2011}. 
There is a nearby molecular peak MF10 at the western side of the submillimeter emission peak \citep{favre2011}. 
Because the position offset between MF10 and Source~I is significant and, in addition, there is no emission from organic molecules at Source~I \citep{beuther2005}, they are unlikely to be physically associated. 

By using previous observational results of interferometric continuum observations at centimeter, millimeter, and submillimeter wavelengths, we plot a SED of Source~I as shown in Figures \ref{fig-sed} and Table \ref{tab-flux} \citep[see][and references therein]{plambeck2013}. 
At several frequencies, there are more than two observational results showing different flux densities. 
For example, flux density of our band~6 data, 284~mJy, is consistent with that of CARMA, 310~mJy \citep{plambeck2013}, with in the errors $\sim$10\%. 
On the other hand, our band~7 data is intermediate between those of \citet{beuther2004} and \citet{tang2010} observed with SMA. 
The reason for such discrepancies could be that insufficient uv coverage affects the flux density measurements due to strong negative sidelobes from the bright extended emission in the complex structure of the Orion~KL region as suggested in \citet{plambeck2013}. 
Another possibility could be a different degree of missing fluxes due to the different uv coverages. 
To avoid such artifact, we employ the flux density observed with the highest resolution observations at each frequency. 
The data employed in the following discussion are indicated by Y in Table \ref{tab-flux}. 

\subsubsection{H$^{-}$ free-free emission}

As shown in Figure \ref{fig-sed} and Table \ref{tab-sed}, the SED of Source~I can be fitted by a single power law function 
\begin{eqnarray}
\log F_{\nu}^{\rm{pl}} & = & p + q \log \nu
\end{eqnarray}
with the power law index $q$ of 1.97$\pm$0.10 for the frequency range between 8.4~GHz and 690~GHz. 
The power law index is consistent with that expected for the black body radiation, $q$=2.0, suggesting that the emission could be optically thick even at the submillimeter wavelength. 
As proposed for typical massive YSOs, the emission mechanism of the centimeter to submillimeter continuum emission from Source~I is a combination of graybody radiation from dust and free-free radiation from fully ionized gas \citep{beuther2004,beuther2006}. 
However, the free-free emission from such a compact H{\sc{ii}} region is ruled out as a main opacity source of continuum emissions according to the observed source size and brightness temperature of $\sim$1500~K at 43~GHz \citep{reid2007, plambeck2013}. 
Observed SED could be reconciled by the free-free emission only if the beam filling factor (source size and/or clumpiness) were much smaller than 1 or the gas temperature were much lower than that expected for the fully ionized gas of $\sim$8000~K. 
Alternatively, it has been suggested that an H$^{-}$ free-free radiation emitted from neutral gas could explain the single power law SED \citep{reid2007, plambeck2013}. 

In order to explain observed SED of Source~I from centimeter to submillimeter wavelength, we first evaluate physical properties of Source~I in the case of the H$^{-}$ free-free radiation. 
Detailed derivation of the H$^{-}$ free-free radiation is summarized in Appendix following the discussion by \citet{reid1997}. 
Hereafter we assume that Source~I is an edge-on disk with uniform density and temperature for simplicity. 
The apparent size of Source~I of 0.20\arcsec$\times$0.03\arcsec \ is employed \citep{plambeck2013}, corresponding to the disk diameter and thickness of 84~AU and 12.6~AU, respectively. 
With these parameters, a total hydrogen density of 10$^{12}$~cm$^{-3}$ corresponds to a column density of 1.26$\times$10$^{27}$~cm$^{-2}$ (with a diameter of 84~AU) and a total mass of 0.20$M_{\odot}$. 

As suggested from Figure \ref{fig-sed}, Source~I seems to be optically thick at 245-339~GHz or even at 690~GHz. 
Thus, the turnover frequency at which optical depth of the H$^{-}$ free-free radiation becomes unity is as high as $>$300-600~GHz (see Appendix). 
Using the relationship between the total hydrogen density (a sum of molecular and atomic hydrogen) and temperature to explain the turnover frequency (Figure \ref{fig-turnfreq2} and Table \ref{tab-turnfreq} in Appendix), we can constrain the range of gas density and temperature. 

According to the SMA continuum observations \citep{tang2010}, the mass of the main dust ridge including Source~I is estimated to be 2-12$M_{\odot}$. 
\citet{favre2011} also derive a consistent value of 5.8$M_{\odot}$ for their continuum source Ca, in which Source~I is located at the western edge of the core. 
The total mass of Source~I itself is estimated from the velocity structure of the rotating disk traced by the SiO masers and submillimeter H$_{2}$O lines \citep{kim2008, matthews2010, hirota2014a}, $>$7$M_{\odot}$, infrared spectroscopy \citep{testi2010}, 10$M_{\odot}$, or a momentum conservation law suggested by proper motion measurement \citep{goddi2011, bally2011}, 20$M_{\odot}$. 
Thus, the circumstellar disk mass of Source~I would be much lower than that of the central protostar and its host core, $\ll$10$M_{\odot}$. 
With this constraint, the total hydrogen density is lower than 5$\times$10$^{13}$~cm$^{-3}$ as indicated in Figure \ref{fig-turnfreq2} in Appendix. 
In this case, the gas temperature would be larger than 1200~K for the turnover frequency of $>$300~GHz 
 (Figure \ref{fig-turnfreq2} and Table \ref{tab-turnfreq} in Appendix). 
If the gas temperature is about 3000~K or higher, the lower gas density of $\sim$10$^{11}$~cm$^{-3}$ is allowed to explain the turnover frequency of $>$300~GHz. 

Using the calculated optical depth, $\tau_{\nu,\rm{H}}$ and $\tau_{\nu,\rm{H}_{2}}$ for the above temperature (1200-3000~K) and density (10$^{11}$-10$^{14}$~cm$^{-3}$) ranges, we calculate the flux density $F_{\nu}$ at each frequency for the source size of $\Omega$; 
\begin{eqnarray}
F_{\nu} & = & \int \frac{2 \nu^{2} kT_{B}}{c^{2}} d\Omega \nonumber \\
          & =  & \int \frac{2 \nu^{2} kT}{c^{2}} (1-e^{-\tau_{\rm{total}}}) d\Omega  \\
T_{B} & = & (1-e^{-\tau_{\rm{total}}} ) T \\
\tau_{\rm{total}} & = & \tau_{\nu,\rm{H}} + \tau_{\nu,\rm{H}_{2}} 
\end{eqnarray}
The results are shown in Figure \ref{fig-flux}. 
Higher temperature or higher density results can explain the single power law SED even for the SMA result at 690~GHz. 
In the case of total hydrogen densities of 10$^{11}$~cm$^{-3}$ and 10$^{12}$~cm$^{-3}$ corresponding to the mass of $\sim$0.02$M_{\odot}$-0.2$M_{\odot}$ (Figures \ref{fig-flux}(a) and (b)), gas temperatures of higher than 2700~K and 1800~K, respectively, can explain the SED of Source~I with single power law index of 2.0 up to 339~GHz band. 
For larger densities of  an order of 10$^{13}$~cm$^{-3}$ or the mass of a few $M_{\odot}$ (Figure \ref{fig-flux}(c)), the lower temperature of 1500~K can also match the slope of the SED of Source~I up to 339~GHz. 
On the other hand, the higher temperature results tend to overestimate the flux densities by a factor of $\sim$2. 
When the emission is optically thick, the flux density is proportional to the apparent source size and gas temperature. 
To account for the observed flux density of Source~I, the lower gas temperature $\sim$1500~K would be more plausible if the source is resolved with the beam size of ALMA. 
Alternatively, the angular size of Source~I, $\Omega$, would have smaller by a factor of 2 or the source structure would be clumpy with the beam filling factor of 50\%. 
Higher resolution observations to resolve the source structure with accurate flux calibration will be able to solve this degeneracy of source size and temperature. 

\citet{hirota2014a} reported that the vibrationally excited H$_{2}$O line at 336~GHz is emitted from a circumstellar disk. 
The 336~GHz H$_{2}$O line is thought to be excited thermally with an excitation temperature of $>$3000~K probably heated via an accretion shock. 
This excitation temperature is slightly higher than that estimated from the present continuum data. 
According to the velocity structure, the 336~GHz H$_{2}$O line map is interpreted as a ring-like structure with a diameter 0.2\arcsec \ (84~AU), although the spatial resolution is still insufficient to resolve the structure. 
It is proposed that the ring-like structure would reflect the distribution of the hot molecular (H$_{2}$O) gas probably heated via accretion \citep{hirota2014a}. 
Alternatively, it is also likely that the 336~GHz H$_{2}$O line could trace only the edge of the disk due to the high continuum opacity even at 336~GHz as suggested by the SED (Figure \ref{fig-sed}). 
If the optically thick H$^{-}$ free-free radiation is the main source of opacity at 336~GHz, the latter scenario would be more plausible. 

\subsubsection{H$^{-}$ free-free emission + thermal dust emission}

\begin{figure}
\begin{center}
\includegraphics[width=7.5cm]{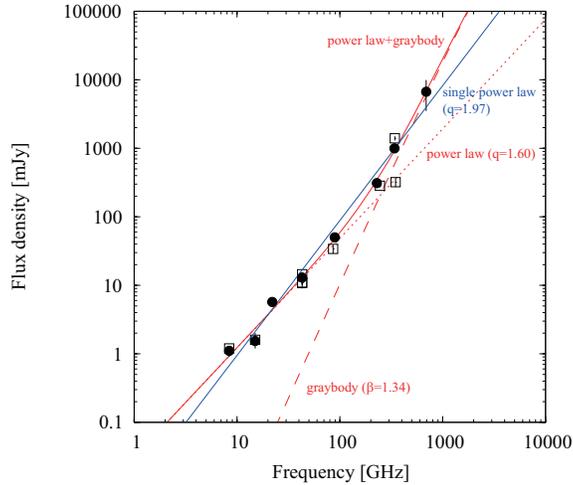}
\caption{
Spectral energy distribution (SED) of Source~I.
A solid blue line indicates the best fit single power law model, $F_{\nu}=p \nu^{q}$ with the index of $q=1.97\pm0.10$. 
A solid red line indicates the combination of power law and graybody SED. 
For the power law and graybody terms are shown by the red dotted and dashed lines, respectively. 
Open squares represent all the data including the present ALMA results \citep[][and references therein]{plambeck2013}. 
Filled circles represent the data employed in the fitting, as indicated Y in Table \ref{tab-flux}. 
}
\label{fig-sed}
\end{center}
\end{figure}

\begin{figure*}
\begin{center}
\includegraphics[width=15cm]{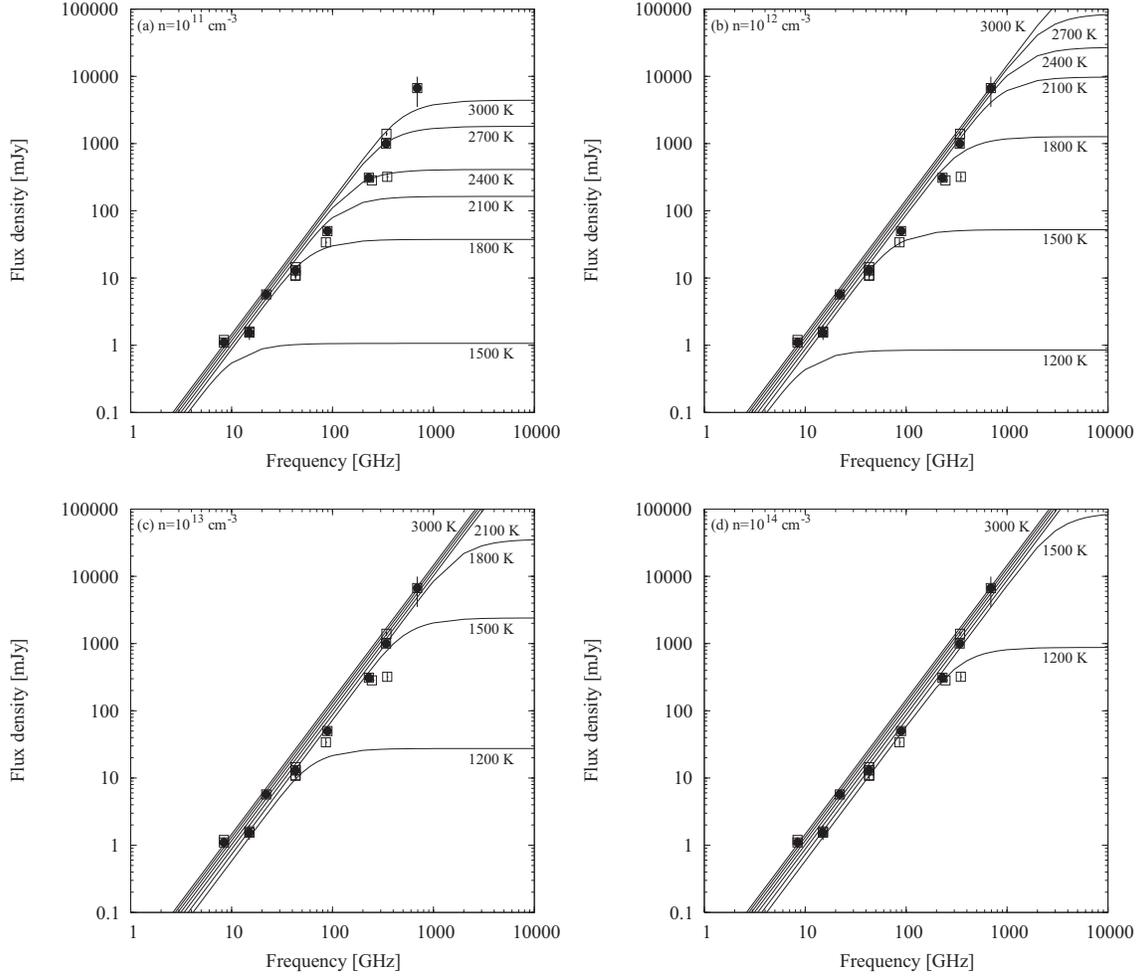}
\caption{
SED of Source~I predicted from H$^{-}$ free-free emission with total hydrogen densities, $n$(H)+2$n$(H$_{2}$) of (a) 10$^{11}$~cm$^{-3}$, (b) 10$^{12}$~cm$^{-3}$, (c) 10$^{13}$~cm$^{-3}$, and (d) 10$^{14}$~cm$^{-3}$.
The path length is assumed to be 84~AU, which is equal to the major axis of Source~I (0.2\arcsec). 
Calculations are done with temperatures from 1200~K to 3000~K with a step of 300~K as shown by solid lines. 
Filled and open squares are the same as Figure \ref{fig-sed}. 
Note that the SED at the temperature of 1200~K with a total hydrogen density of 10$^{11}$~cm$^{-3}$ (a) is not plotted because of the lower turnover frequency than the plotting range. 
}
\label{fig-flux}
\end{center}
\end{figure*}

In Figure \ref{fig-sed}, our ALMA band~7 data still suggest a marginal excess flux compared with the single power law fitting result. 
The flux density at 690~GHz also shows a significant excess probably originated from thermal dust emission as suggested by \citet{beuther2006} and \citet{plambeck2013}. 
There seems to be systematic deviation between the best-fit model and observed results from 43~GHz and 229~GHz, suggesting a change in the emission mechanism at the middle of the centimeter to millimeter wavelength. 
Thus, we next consider the contribution from thermal graybody emission from dust. 
The H$^{-}$ free-free emission and dust graybody emission contribute to the flux densities at lower and higher frequency regions, respectively. 
In this case, the flux density $F_{\nu}$ at each frequency $\nu$ is fitted to the sum of power law function due to the H$^{-}$ free-free radiation and the graybody radiation from the thermal dust emission; 
\begin{eqnarray}
F_{\nu} & = & F_{\nu}^{\rm{pl}} + F_{\nu}^{\rm{gb}}. 
\end{eqnarray}

The combined fitting result gives the lower power law index of 1.60$\pm$0.24, than that of the single-power law fitting, 1.97 (Table \ref{tab-sed}). 
The significant amount of the observed flux densities at higher frequency ($>$200~GHz) can be explained by the graybody radiation rather than the H$^{-}$ free-free radiation. 
In such a case, the turnover frequency could be as low as 200~GHz. 
The required density or temperature under this condition are reduced by a factor of about $(2/3)^{2}$ when we change the turnover frequency from 300~GHz to 200~GHz (see equation (\ref{eq-turnfreq})). 

Because of the lack of far-infrared wavelength data corresponding to the flux maximum of the graybody function, we cannot constrain the dust temperature of the emitting region. 
It is possible that the dust graybody emission arises from the different volume of gas mainly in a cooler outer layer of Source~I or the same volume of gas as the H$^{-}$ free-free emission at the temperature higher than $\sim$1000~K. 
In the former case, the total dust+gas mass contributed by the graybody SED is estimated to be 0.082~M$_{\odot}$ assuming the dust to gas mass ratio of 100 and the dust temperature of 100~K (Table \ref{tab-sed}). 
This value is smaller than that obtained from the SMA observations, 0.2~M$_{\odot}$ \citep{beuther2004}. 
In the latter case, the dust temperature would be higher than 1000~K as estimated from the continuum emission \citep{reid2007, plambeck2013} and vibrationally excited H$_{2}$O lines \citep{hirota2014a}. 
The resultant circumstellar dust+gas mass would be reduced by a factor of 10, $\sim$0.008~M$_{\odot}$, compared with that of 100~K. 
Thus, the dust mass would be smaller by a factor of 100 (i.e. assumed dust to gas mass ratio), 8$\times$10$^{-5}$~M$_{\odot}$. 
This value is much smaller than the gas mass estimated from the H$^{-}$ opacity at this temperature as discussed above (see Figure \ref{fig-turnfreq2}). 
In either case, our result may suggest that the dust to gas mass ratio is unusually low in the close vicinity of Source~I within $\sim$100 AU in diameter because of the dust sublimation under the temperature of $>$1000~K \citep[e.g.][]{reid2007}. 

A power law index smaller than 2.0 would reflect a density structure of the emitting region \citep[e.g. discussion in ][]{beuther2004,plambeck2013}. 
In the present analysis, we do not make any correction for above flux densities such as the different source/beam sizes, which could more or less induce uncertainties in the derived parameters. 
To better constrain the emission mechanism, higher resolution observations with accurate flux calibration are required to resolve the size of Source~I. 

\begin{deluxetable}{rccclcl}
\tablewidth{0pt}
\tabletypesize{\scriptsize}
\tablecaption{Flux densities of Source~I
 \label{tab-flux}}
\tablehead{
\colhead{Frequency} & \colhead{Flux}  & \colhead{HPBW}  & \colhead{}      & \colhead{}      & \colhead{}      &\colhead{}    \\
\colhead{[GHz]}     & \colhead{[mJy]} & \colhead{[mas]} & \colhead{Array} & \colhead{Epoch} & \colhead{SED\tablenotemark{a}} & \colhead{Reference} 
}
\startdata
  8.4    &     1.1$\pm$0.2  &       220        & VLA    & 1994.3   & Y & \citet{menten1995}         \\
  8.4    &     1.2$\pm$0.1  &  262$\times$217  & VLA    & 2006.4   &   & \citet{gomez2008}          \\
 15 \ \  &    1.54$\pm$0.18 &  140$\times$130  & VLA    & 1986.3   & Y & \citet{felli1993a}         \\
 15 \ \  &     1.6$\pm$0.4  &       150        & VLA    & 1990 \ \ &   & \citet{felli1993b}         \\
 22 \ \  &     5.7$\pm$0.9  &  109$\times$97   & VLA    & 1991.5   & Y & \citet{forbrich2008}       \\
 43 \ \  &      13$\pm$2    &       250        & VLA    & 1994.3   &   & \citet{menten1995}         \\
 43 \ \  &    10.8$\pm$0.6  & 1960$\times$1410 & VLA    & 1994.9   &   & \citet{chandler1997}       \\
 43 \ \  &         13       &    41$\times$28  & VLA    & 2000.9   & Y & \citet{reid2007}           \\
 43 \ \  &    14.5$\pm$0.7  &  170$\times$150  & VLA    & 2007.9   &   & \citet{rodriguez2009}      \\
 43 \ \  &      11$\pm$2    &   58$\times$39   & VLA    & 2009.0   &   & \citet{goddi2011}          \\
 86 \ \  &      34$\pm$5    & 1000$\times$380  & BIMA   & 1995.0   &   & \citet{plambeck1995}       \\
 90 \ \  &      50$\pm$5    &  400$\times$350  & CARMA  & 2009.0   & Y & \citet{friedel2011}        \\
229 \ \  &     310$\pm$45   &  150$\times$130  & CARMA  & 2009.1   & Y & \citet{plambeck2013}       \\
245 \ \  &      272$\pm$13   & 630$\times$470  & ALMA  & 2012 \ \ &  & Present study \\
339 \ \  &      849$\pm$29   &  460$\times$410 & ALMA & 2012 \ \ & Y & Present study \\
341 \ \  &    1400$\pm$100  &  800$\times$700  & SMA    & 2009.1   &   & \citet{tang2010}           \\
348 \ \  &     320$\pm$48   &  780$\times$650  & SMA    & 2004.1   &   & \citet{beuther2004}        \\
690 \ \  &    6700$\pm$3200 & 1400$\times$900  & SMA    & 2005.1   & Y & \citet{beuther2006}        \\
\enddata
\tablecomments{See \citet{plambeck2013}. }
\tablenotetext{a}{Y indicates the data employed in the SED fitting. }
\end{deluxetable}

\begin{deluxetable}{lcccc}
\tablewidth{0pt}
\tabletypesize{\scriptsize}
\tablecaption{SED of Source~I
 \label{tab-sed}}
\tablehead{
\colhead{Model}        & \colhead{$p$} & \colhead{$q$} & \colhead{$M$(100~K)[$M_{\odot}$]}  & \colhead{$\beta$}  }
\startdata
Single power law       &  -1.99(19)   & 1.97(10)      & \nodata                            & \nodata        \\                                            
power law$+$graybody &  -1.51(30)   & 1.60(24)      & 0.082(49)                    &  1.34(83) \\
\enddata
\tablecomments{Dust temperature of 100~K is assumed. \\
Numbers in parenthesis represent fitting errors (1$\sigma$) in unit of the last significant digits. }
\end{deluxetable}

\subsection{Hot Core (HKKH5)}

Hot Core in Orion~KL has been recognized as a prototypical dense and hot molecular gas clump which are often found in massive star-forming regions. 
Hot Core is known to show wealth of molecular line emissions in millimeter and submillimeter wavelengths \citep[e.g.][]{beuther2005}. 
A submillimeter continuum source SMM3 detected with SMA \citep{zapata2011} corresponds to this compact millimeter/submillimeter source. 
This source corresponds to the 3~mm continuum emission C18 \citep{friedel2011} while it cannot be seen in the molecular line map of methyl formate \citep{favre2011}. 
It is most likely that methyl formate is deficient in the Hot Core where nitrogen-bearing organic molecules are dominant. 
According to the recent molecular line observations, Orion Hot Core is most likely heated externally by the explosive outflow rather than by an embedded self-luminous source \citep{zapata2011}. 
The mass derived from our ALMA data is 0.159$M_{\odot}$ at the largest estimate assuming the temperature of 20~K. 
The dust opacity index derived from the graybody SED model is consistent with 1.0. 
As discussed for Source~I, the mass of the main dust ridge including Source~I and Hot Core is estimated to be 2-12$M_{\odot}$ \citep{tang2010} and 5.8$M_{\odot}$ \citep{favre2011}. 
Although the above estimated mass could not resolve more compact structures such as Source~I and SMA1, our result only recovers a part of the total mass of the dense gas. 

\subsection{SMA1 (HKKH7)}

A submillimeter continuum source SMA1 is first identified by \citet{beuther2004} in the SMA observations at 348~GHz. 
This source is identified to be the 3~mm continuum source C21 \citep{friedel2011} and the methyl formate peak MF6 \citep{favre2011}. 
SMA1 is proposed to be a powering source of the explosive outflow \citep{beuther2008}, although the origin of this explosive outflow is still under debate \citep[e.g.][]{zapata2009,bally2011,goddi2011}.
The flux density and peak intensity of our ALMA band~7 data are 534~mJy and 237~mJy~beam$^{-1}$, respectively. 
On the other hand, SMA1 was not resolved with the previous SMA observations with the flux density and peak intensity of 360~mJy and 360~mJy~beam$^{-1}$, respectively \citep{beuther2004}. 
The difference between SMA and ALMA observations could be attributed to the different uv coverage. 
The dust opacity index, $\beta$ is consistent with 1.0. 
The derived circumstellar mass of SMA1 is only $\sim$0.18$M_{\odot}$ assuming the dust temperature of 20~K. 
It is much smaller than the total mass of the main dust ridge of this region as discussed above. 

\subsection{IRc7 (HKKH6) and IRc4 (HKKH10)}

We clearly see an anti-correlation between the millimeter/submillimeter emission and the mid-infrared emission (Figure \ref{fig-label}). 
However, we detect compact continuum emission sources associated with infrared sources IRc4 and IRc7. 
The nature of these sources whether they are heated internally or externally are still unclear. 
The infrared spectrum of IRc4 observed with the Subaru telescope can be fitted to the Planck function with a single temperature of 140~K \citep{okumura2011}, and it is thought to be heated by an external energy source(s) \citep{okumura2011}. 
On the other hand, a color temperature map obtained from longer wavelength observations with SOFIA suggests that IRc4 is a self-luminous source \citep{debuizer2012}. 
For IRc7, it is proposed that an outflow or radiation from Source~n would form a fan-like structure in the infrared emission \citep{greenhill2004}. 
According to near-infrared polarization observations with Hubble Space Telescope (HST), IRc4 is thought to be a reflection nebula illuminated by a nearby source(s) while IRc7 is most likely heated by an embedded YSO \citep{simpson2006}. 
In the 3~mm continuum map, IRc7 is identified as a compact source C23 but IRc4 is only found as a diffuse emission \citep{friedel2011}. 
We detect compact high density cores with the sizes and H$_{2}$ densities of $\sim$200-300~AU and $>$10$^{8}$~cm$^{-3}$ at the positions of IRc4 and IRc7. 
Circumstellar masses are estimated to be larger than 0.145$M_{\odot}$ and 0.086$M_{\odot}$ for IRc4 and IRc7, respectively. 

Except for IRc4, IRc7, and BN as discussed below, we cannot detect millimeter/submillimeter counterparts at the position of other IRc sources. 

\subsection{Compact Ridge (HKKH11)}

We identify a compact millimeter/submillimeter emission corresponding to a submillimeter source SMM1 \citep{zapata2011}, 3~mm continuum source C32 \citep{friedel2011} and methyl formate peak MF1 \citep{favre2011} in the Compact Ridge. 
The mass of the continuum source at the Compact Ridge is estimated to be 4$M_{\odot}$ \citep{tang2010}. 
\citet{eisner2008} identified a compact 1.3~mm continuum source named HC~438 in the Compact Ridge by using CARMA at the highest resolution of 0.5\arcsec \ compared with other previous results. 
They detected the 1.3~mm continuum emission with a flux density of 67.8$\pm$14.2~mJy. 
This is consistent with our ALMA band~6 data at a similar spatial resolution. 
The circumstellar mass of HC~438 is derived to be 0.085~$M_{\odot}$ with the assumed dust temperature of 20~K \citep{eisner2008}. 
Our result, 0.105$M_{\odot}$ agrees well with that of \citet{eisner2008}. 
If the higher temperature of 100~K derived from the methyl formate data \citep{favre2011} is employed, the smaller mass of 0.010$M_{\odot}$ is obtained. 

As discussed by \citet{favre2011}, there are possible counterpart physically associated with the millimeter/submillimeter continuum source; a radio continuum source labeled R \citep{felli1993a}, a molecular peak traced by methyl formate lines, MF1 \citep{favre2011}, and optical source Parenago~1822. 
In addition, an extremely bright  22~GHz H$_{2}$O maser source appears in the Compact Ridge at the systemic velocity of $\sim$ 6.9~km~s$^{-1}$ and 7.5~km~s$^{-1}$. 
It is sometimes called supermaser \citep[][and references therein]{hirota2011, hirota2014b}. 
The supermaser is located within the millimeter/submillimeter continuum source detected with ALMA. 

\citet{favre2011} suggest that the supermaser would be related to the shocked molecular gas of the Compact Ridge \citep{liu2002}. 
It is also suggested that the maser burst could occur as a result of an interaction with outflows and a pre-existing YSO in the Compact Ridge \citep{garay1989}. 
\citet{hirota2014b} speculate that the compact continuum source would play an important role to amplify the supermaser which can supply plenty of ambient molecular gas. 
Near the supermaser, a cluster of other 22~GHz H$_{2}$O maser features detected with VLA \citep{gaume1998} are distributed around the continuum source. 
It may imply an embedded YSO as a powering source of the H$_{2}$O maser cluster \citep{hirota2014b}. 

We note that an infrared source IRc5 is located at 1\arcsec east of the ALMA continuum peak position and is corresponding to another 3~mm continuum peak, C29 \citep{friedel2011}. 
The infrared spectrum of IRc5 can be fitted with the Planck function at the temperature of 120~K \citep{okumura2011}. 
Similar to IRc4, IRc5 is thought to be a shocked molecular gas heated externally \citep{okumura2011}. 
Thus, IRc5 is not physically associated with the millimeter/submillimeter compact source detected with ALMA. 

\subsection{BN}

Orion BN object is another interesting source regarding the energy source of this region \citep{galvan-madrid2012, plambeck2013} and a possible counterpart of the close encounter with Source~I \citep{gomez2008,goddi2011, bally2011} or $\theta^{1}C$ \citep{chatterjee2012}. 
It is a massive YSO associated with a hypercompact H{\sc{ii}} region ionized by a zero age main sequence B star \citep{plambeck2013}. 
According to the ALMA SV data \citep{galvan-madrid2012} with a spatial resolution of 1.32\arcsec$\times$0.62\arcsec, a compact continuum emission at  230~GHz is detected with a flux density of 126~mJy by using the baselines longer than 100~k$\lambda$ to filter-out extended emission. 
On the other hand, the higher resolution CARMA observation detects the 229~GHz continuum emission of 240~mJy with the resolution of 0.14\arcsec \ \citep{plambeck2013}. 
A continuum emission source associated with the BN object is marginally seen in our ALMA bands~6 and 7 images outside the primary beam size of the ALMA antenna (Figures \ref{fig-mapb6} and \ref{fig-mapb7}). 
Flux densities at bands~6 and 7 are derived to be 55$\pm$4~mJy and 33$\pm$6mJy, respectively. 
When we correct for an effect of the primary beam attenuation by a factor of 2, the calibrated peak intensities are 110~mJy and 66~mJy, respectively. 
Marginal detection in our ALMA image at band~6 does not contradict to the results of \citet{galvan-madrid2012} although our ALMA results are quite uncertain. 
Thus, we do not consider the BN object as an identified source in the present paper. 

\subsection{Source~n}

Source~n is also detected in the SMA observation at 348~GHz \citep{beuther2004}. 
It is proposed to be a YSO associated with a bipolar radio jet \citep{menten1995} and a circumstellar disk traced by a mid-infrared emission \citep{greenhill2004}. 
Our ALMA images at bands~6 and 7 show no compact emission at the position of Source~n. 
The upper limit ($5\sigma$) of their peak intensities are 25~mJy~beam$^{-1}$ for both bands~6 and 7. 
The peak intensity derived from the SMA observation, 300~mJy~beam$^{-1}$ is much larger than our ALMA result. 
The non-detection of Source~n in our ALMA images are possibly due to our lower sensitivities for extended emission components. 

\section{Summary}

We have carried out millimeter/submillimeter continuum imaging of the Orion~KL region by using newly constructed ALMA at a resolution of 0.5\arcsec, \ corresponding to a linear scale of 200~AU. 
Compact continuum emission sources are detected and 11 sources are identified at both at band~6 (245~GHz) and band~7 (339~GHz). 
They include some of remarkable sources in Orion~KL such as  Source~I, Hot Core, SMA1, IRc4, IRc7, and Compact Ridge.  
Their physical properties such as size, mass, H$_{2}$ number density and column densities are discussed by employing published 3~mm continuum data \citep{friedel2011} to construct SED of dust graybody emission. 

Among these identified sources, SED of Source~I, which is a dominant energy source in Orion~KL, are extensively studied by using previous observational results from centimeter to submillimeter wavelengths \citep[see references in ][]{plambeck2013}. 
The SED model with H$^{-}$ free-free emission is presented following the discussion by \citet{reid1997} to explain the power-law index of the SED, 1.97,  consistent with an optically thick emission. 
We introduce a turnover frequency of the H$^{-}$ free-free emission to constrain the gas temperature and total hydrogen density of the circumstellar disk of Source~I. 
As a result, total hydrogen density of 10$^{11}$-10$^{14}$~cm$^{-3}$ are required to account for the SED with a single power-law index of 2.0 under the temperature of 1200-3000~K in the case of turnover frequency of $\sim$300~GHz. 
When we employ the combination of dust graybody emission and power-law SED, the turnover frequency would be as low as 200~GHz, which reduces the estimated temperature and/or density by a factor of $(2/3)^{2}$. 
The fitting result yields a smaller power-law index of 1.60, suggesting a compact size or clumpy structure of the emission region unresolved with the present study \citep{beuther2004,plambeck2013}. 
The estimated temperature, density and source size are strongly coupled with each other and hence, future higher resolution observations with ALMA will be key issues to solve these degeneracy in physical properties. 

\acknowledgements
We are grateful to S. Okumura for providing a Subaru mid-infrared image. 
This paper makes use of the following ALMA data: ADS/JAO.ALMA\#2011.0.00009.SV and 2011.0.00199.S. 
ALMA is a partnership of ESO (representing its member states), NSF (USA) and NINS (Japan), together with NRC (Canada) and NSC and ASIAA (Taiwan), in cooperation with the Republic of Chile. 
The Joint ALMA Observatory is operated by ESO, AUI/NRAO and NAOJ. 
We thank the staff at ALMA for making observations and reducing the science verification data. 
T.H. is supported by the MEXT/JSPS KAKENHI Grant Numbers 21224002, 24684011, and 25108005, and the ALMA Japan Research Grant of NAOJ Chile Observatory, NAOJ-ALMA-0006. 
M.H. is supported by the MEXT/JSPS KAKENHI Grant Numbers 24540242 and 25120007.

{\it Facilities:} \facility{ALMA}.

\appendix
\section{H$^{-}$ free-free radiation}

The free-free transition of electrons interacting with neutral hydrogen atoms and molecules is called H$^{-}$ free-free emission. 
The formulas for the absorption coefficients of H$^{-}$ free-free emission were first developed by \citet{dalgarno1966}. 
Then, \citet{reid1997} applied the H$^{-}$ free-free emission mechanism to explain radio photospheres of Mira-type variables by simplifying the formula of H$^{-}$ free-free absorption coefficients. 
Taking into account a similarity in observed properties between Mira-type variables and Source~I, the H$^{-}$ free-free emission can be applied to Source~I \citep{beuther2004, reid2007, plambeck2013}. 
Here we follow the discussion by \citet{reid1997} to construct a SED of Source~I, assuming the source diameter (path length) of 84~AU corresponding to the major axis of Source~I of 0.2\arcsec. 

The H$^{-}$ free-free absorption coefficients at 10~GHz for unit electron pressure per hydrogen atom, $\kappa_{\rm{10 GHz,H}}$, and molecule, $\kappa_{\rm{10 GHz,H}_{2}}$, at the temperature $T$ are approximately expressed as 
\begin{eqnarray}
\kappa_{\rm{10 GHz,H}} & = & a_{0,\rm{H}}+a_{1,\rm{H}}T+a_{2,\rm{H}}T^{2}+a_{3,\rm{H}}T^{3}, \label{eq-kappah}\\
a_{0,\rm{H}} & = & +3.376 \times 10^{-17}, \\
a_{1,\rm{H}} & = & -2.149 \times 10^{-20}, \\
a_{2,\rm{H}} & = & +6.646 \times 10^{-24}, \\
a_{3,\rm{H}} & = & -7.853 \times 10^{-28},
\end{eqnarray}
and
\begin{eqnarray}
\kappa_{\rm{10 GHz,H}_{2}} & = & a_{0,\rm{H}_{2}}+a_{1,\rm{H}_{2}}T+a_{2,\rm{H}_{2}}T^{2}+a_{3,\rm{H}_{2}}T^{3}, \label{eq-kappah2}\\
a_{0,\rm{H}_{2}} & = & +8.939 \times 10^{-18}, \\
a_{1,\rm{H}_{2}} & = & -3.555 \times 10^{-21}, \\
a_{2,\rm{H}_{2}} & = & +1.100 \times 10^{-24}, \\
a_{3,\rm{H}_{2}} & = & -1.319 \times 10^{-28}, 
\end{eqnarray}
respectively, in unit of cm$^{4}$~dyn$^{-1}$\citep{reid1997}.  
The above equations are valid under low temperature ($<$3000~K) and in long wavelength ($>$3~$\mu$m) as discussed in \citet{reid1997}. 
Because the absorption coefficient is proportional to the inverse square of the frequency, $\nu$, then the absorption coefficients can be written as 
\begin{eqnarray}
\alpha_{\nu,\rm{H}} & = & \alpha_{\rm{10~GHz,H}} \left( \frac{\rm{10~GHz}}{\nu \ \rm{GHz}} \right)^2 = \kappa_{\rm{10~GHz,H}} n_{e} n_{\rm{H}} k T \left( \frac{\rm{10~GHz}}{\nu \ \rm{GHz}} \right)^2 \\
\alpha_{\nu,\rm{H}_{2}} & = & \alpha_{\rm{10~GHz,H}} \left( \frac{\rm{10~GHz}}{\nu \ \rm{GHz}} \right)^2 = \kappa_{\rm{10~GHz,H}_{2}} n_{e} n_{\rm{H}_{2}} k T \left( \frac{10~\rm{GHz}}{\nu \ \rm{GHz}} \right)^2 
\end{eqnarray}
where $k$ is the Boltzmann constant and $n_{e}$, $n_{\rm{H}}$, and $n_{\rm{H}_{2}}$ are number densities of electron, hydrogen atom, and hydrogen molecule, respectively. 
The optical depths for H$^{-}$ free-free emission for hydrogen atoms and molecules are 
\begin{eqnarray}
\tau_{\nu,\rm{total}} & = & \tau_{\nu,\rm{H}} + \tau_{\nu,\rm{H}_{2}} \label{eq-tautotal} \\
\tau_{\nu,\rm{H}} & = & \alpha_{\nu,\rm{H}} L  =  \kappa_{\rm{10~GHz,H}} n_{e} n_{\rm{H}} L k T \left( \frac{\rm{10~GHz}}{\nu \ \rm{GHz}} \right)^2 \label{eq-tauh} \\
\tau_{\nu,\rm{H}_{2}} & = & \alpha_{\nu,\rm{H}} L = \kappa_{\rm{10~GHz,H}_{2}} n_{e} n_{\rm{H}_{2}} L k T \left( \frac{\rm{10~GHz}}{\nu \ \rm{GHz}} \right)^2 \label{eq-tauh2} 
\end{eqnarray}
where $L$ represents the path length along a line of sight. 

The optical depths, (\ref{eq-tauh}) and (\ref{eq-tauh2}), are obtained by solving the thermal equilibrium of hydrogen molecules, atoms, and ions at the fixed temperature \citep{cox2000}; 
\begin{eqnarray}
\frac{n_{\rm{H}}^2}{n_{\rm{H}_{2}}} & = & \left( \frac{2 \pi m_{\rm{H}_{2}} k T}{h^2} \right)^{\frac{3}{2}} \exp \left( -\frac{D}{k T}\right) \frac{U_{\rm{H}}(T)^2}{U_{\rm{H}_{2}}(T)}, \\
\frac{n_{\rm{H}^{+}}}{n_{\rm{H}}} n_{e} & = & \left( \frac{2 \pi m_{e} k T}{h^2} \right)^{\frac{3}{2}} \exp \left( -\frac{I}{k T}\right) \frac{2 U_{\rm{H}^{+}}(T)}{U_{\rm{H}}(T)}. 
\label{eq-sahaH}
\end{eqnarray}

Here, $m_{e}$ and $m_{\rm{H}}$ are the mass of electron and hydrogen atom, respectively, and $m_{\rm{H}_{2}}$ is a reduced mass of the hydrogen molecule, 0.504 atomic mass unit \citep{cox2000}. 
$D$ and $I$ are dissociation energy of hydrogen molecule, 4.478~eV, and ionization energy of hydrogen atom, 13.598~eV, respectively \citep{cox2000}. 
Partition function of the hydrogen ion and atom are denoted as $U_{\rm{H}^{+}}(T)$=1 and $U_{\rm{H}}(T)$=2, respectively, 
The partition function of the hydrogen molecule is calculated as $kT/hcB$ if $kT/hcB$ is much larger than 1 with the rotational constant $B$ in unit of cm$^{-1}$. 
For the hydrogen molecule, $B$ is 60.85~cm$^{-1}$ \citep{cox2000} and hence, this approximation is valid in the temperature of $\gg$88~K. 

\begin{figure}
\begin{center}
\includegraphics[width=15cm]{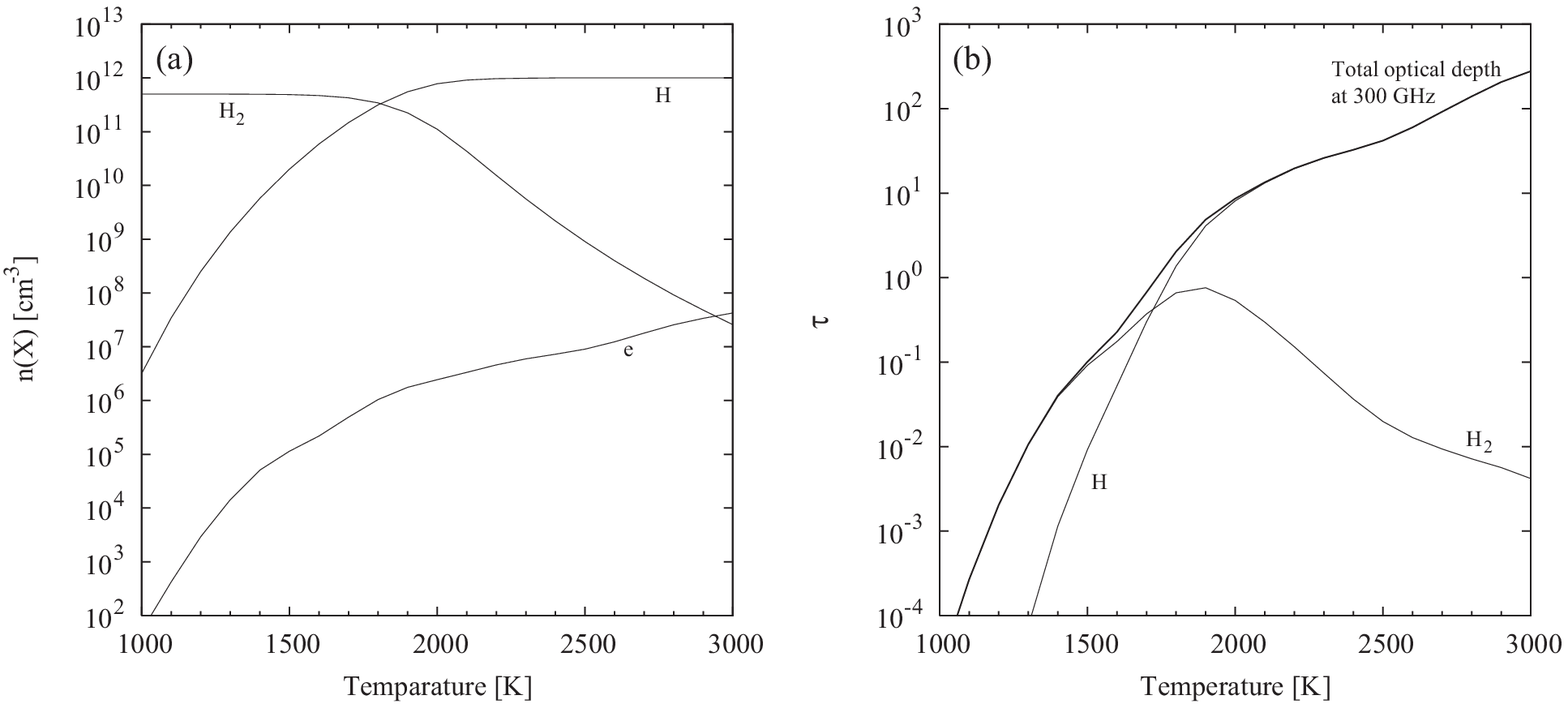}
\caption{
(a) Calculated number density of hydrogen atom, hydrogen molecule, and electron 
with the assumed total hydrogen density of $n$(H)+2$n$(H$_{2}$)=10$^{12}$~cm$^{-3}$. 
Electron density is calculated from the Saha equations of Na, Mg, Al, Si, K, Ca, Fe, and H (see text). 
(b) Calculated optical depth of H$^{-}$ free-free emission at 300~GHz. 
Contributions from H atom and H$_{2}$ molecule are plotted separately. 
The total hydrogen density of $n$(H)+2$n$(H$_{2}$)=10$^{12}$~cm$^{-3}$ and path length of 84~AU are assumed. 
}
\label{fig-saha}
\end{center}
\end{figure}

As discussed in \citet{reid1997}, main sources of electron under the temperature lower than $\sim$4000~K are abundant metals such as Na, Mg, Al, Si, K, Ca, and Fe while hydrogen ion is a dominant source of electron at the temperature higher than $\sim$4000~K. 
We evaluate the ionization degree of Na, Mg, Al, Si, K, Ca, and Fe, along with H, by solving their Saha equations;  
\begin{eqnarray}
\frac{n_{i}}{n_{0}} n_{e} & = & \left( \frac{2 \pi m_{e} k T}{h^2} \right)^{\frac{3}{2}} \exp \left( -\frac{I}{k T}\right) \frac{2 U_{i}(T)}{U_{0}(T)}, 
\end{eqnarray}
assuming the Solar-system chemical abundances \citep{cox2000}. 
The ionization energy and partition function for each species are taken from \citet{cox2000} and \citet{gray2005}, respectively. 
First, the Saha equation of each species is solved individually to estimate an ionization degree of each species and a total electron density.  
Then using these initial values, we solve the Saha equations of all species iteratively until all of the solutions are converged. 
In Table \ref{tab-saha}, we tabulate the calculated ionization degree for the temperature range from 1000~K to 3000~K, in which the approximation of the H$^{-}$ free-free absorption coefficients (equations \ref{eq-kappah} and \ref{eq-kappah2}) are valid, with a step of 100~K. 
The results are also shown in Figure \ref{fig-saha}(a). 
Molecular hydrogen is converted to atomic hydrogen at the temperature higher than $\sim$2000~K. 
On the other hand, thermal ionization of hydrogen is not significant until the temperature exceeds $\sim$8000~K. 

Using the number densities of electron, atomic and molecular hydrogen, we calculate the optical depths of the H$^{-}$ free-free radiation. 
Figure \ref{fig-saha}(b) shows the optical depths of the H$^{-}$ free-free radiation as a function of temperature. 
The dominant opacity source of H$^{-}$ free-free emission is molecular hydrogen under the temperature of 1700~K, and the optical depth due to molecular hydrogen shows the maximum at the temperature of 1900~K. 
At the higher temperature range, atomic hydrogen becomes the dominant source of opacity of H$^{-}$ free-free emission. 

The optical depth is proportional to the inverse square of frequency \citep{reid1997}. 
Even if the H$^{-}$ free-free radiation is optically thick at lower temperature, the emission becomes optically thin at higher frequency band. 
Thus, we can define the transition frequency from optically thick to thin regime as a turnover frequency, $\nu_{\rm{to}}$, as employed in the proton-electron free-free radiation. 
At the turnover frequency, the optical depth becomes unity. 
From the equations (\ref{eq-tautotal}), (\ref{eq-tauh}), and (\ref{eq-tauh2}), the turnover frequency can be expressed as 
\begin{eqnarray}
\nu_{\rm{to}} & = & 10 \sqrt{L k T \left( \kappa_{\rm{10 GHz,H}} n_{e} n_{\rm{H}}  + \kappa_{\rm{10 GHz,H}_{2}} n_{e} n_{\rm{H}_{2}}  \right)} \ \rm{GHz}. 
\label{eq-turnfreq}
\end{eqnarray}

\begin{figure}
\begin{center}
\includegraphics[width=15cm]{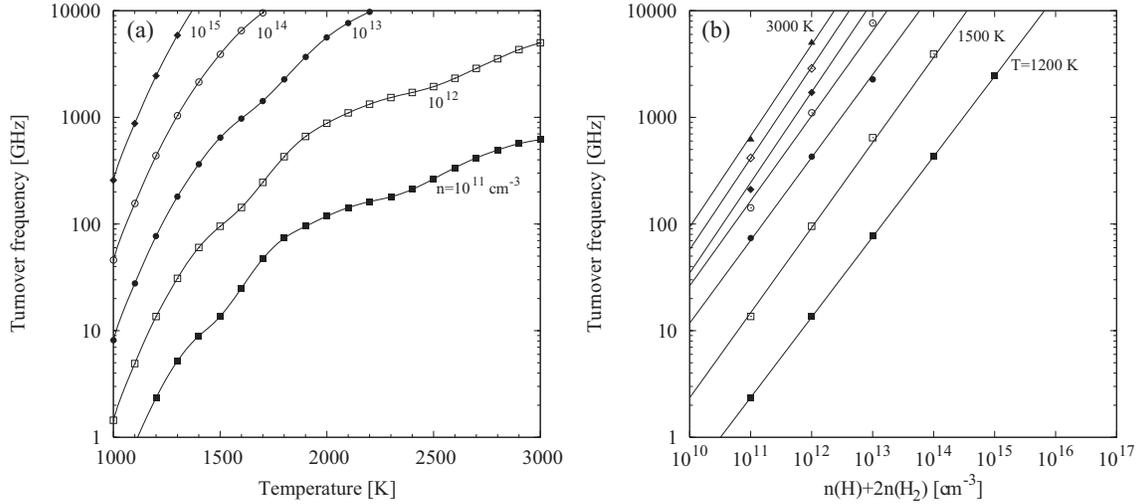}
\caption{
(a) Turnover frequency as a function of temperature for the fixed total hydrogen density from 10$^{11}$~cm$^{-3}$ to 10$^{15}$~cm$^{-3}$. 
(b) Turnover frequency as a function of total hydrogen density for the fixed temperature from 1200~K to 3000~K with a step of 300~K. 
Filled and open symbols represent the calculated values. 
Solid lines represent the best fit power-law function as summarized in Table \ref{tab-turnfreq}. 
The path length is fixed to be 84~AU for both panels. 
}
\label{fig-turnfreq}
\end{center}
\end{figure}

Figures \ref{fig-turnfreq}(a) and \ref{fig-turnfreq}(b) show the turnover frequency as functions of temperature and total hydrogen density, respectively. 
Although the turnover frequency cannot be expressed by a simple formula as a function of temperature, it can be estimated as a simple power law function of the total hydrogen density at a given temperature; 
\begin{eqnarray}
\log \nu_{\rm{to}} & = & A + B \log \bigl[ n({\rm{H}})+2n({\rm{H}}_{2}) \bigr] 
\label{eq-turnfunc} 
\end{eqnarray}

\begin{figure*}
\begin{center}
\includegraphics[width=7.5cm]{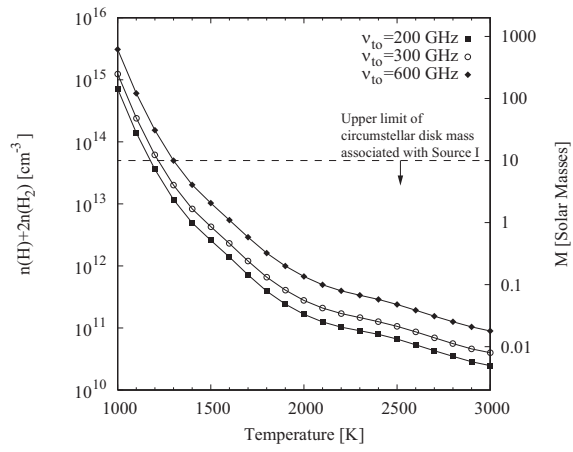}
\caption{
Relationship between temperature and total hydrogen density, $n$(H)+2$n$(H$_{2}$) for the fixed turnover frequency. 
Filled square, open circle, and filled diamond represents the turnover frequency of 200~GHz, 300~GHz, and 600~GHz, respectively. 
The path length is fixed to be 84~AU. 
}
\label{fig-turnfreq2}
\end{center}
\end{figure*}

The fitting results are summarized in Figure \ref{fig-turnfreq}(b) and Table \ref{tab-turnfreq}. 
Using these results, we constrain the gas temperature and the total hydrogen density at the fixed turnover frequency of 200~GHz, 300~GHz, and 600~GHz as shown in Figure \ref{fig-turnfreq2} and Table \ref{tab-turnfreq}. 

\begin{deluxetable}{cccccccccccccccc}
\rotate
\tablewidth{0pt}
\tabletypesize{\scriptsize}
\tablecaption{Fractional abundances of atomic hydrogen, molecular hydrogen and electron
\label{tab-saha}}
\tablehead{
\colhead{} & \colhead{} & \colhead{10$^{11}$~cm$^{-3}$} & \colhead{} & 
\colhead{} & \colhead{} & \colhead{10$^{12}$~cm$^{-3}$} & \colhead{} & 
\colhead{} & \colhead{} & \colhead{10$^{13}$~cm$^{-3}$} & \colhead{} & 
\colhead{} & \colhead{} & \colhead{10$^{14}$~cm$^{-3}$} & \colhead{} \\ 
\cline{2-4} \cline{6-8} \cline{10-12} \cline{14-16} \\
\colhead{$T$(K)} & \colhead{H} & \colhead{H$_{2}$} & \colhead{$e$} & 
\colhead{}          & \colhead{H} & \colhead{H$_{2}$} & \colhead{$e$} & 
\colhead{}          & \colhead{H} & \colhead{H$_{2}$} & \colhead{$e$} & 
\colhead{}          & \colhead{H} & \colhead{H$_{2}$} & \colhead{$e$} \\
}
\startdata
1000 & 1.0(-05) & 5.0(-01) & 2.6(-10) & & 3.2(-06) & 5.0(-01) & 5.2(-11) & & 1.0(-06) & 5.0(-01) & 1.3(-11) & & 3.2(-07) & 5.0(-01) & 3.9(-12) \\ 
1100 & 1.1(-04) & 5.0(-01) & 1.4(-09) & & 3.4(-05) & 5.0(-01) & 4.2(-10) & & 1.1(-05) & 5.0(-01) & 1.3(-10) & & 3.4(-06) & 5.0(-01) & 4.1(-11) \\ 
1200 & 8.0(-04) & 5.0(-01) & 8.9(-09) & & 2.5(-04) & 5.0(-01) & 2.9(-09) & & 8.0(-05) & 5.0(-01) & 9.5(-10) & & 2.5(-05) & 5.0(-01) & 3.0(-10) \\ 
1300 & 4.3(-03) & 5.0(-01) & 4.0(-08) & & 1.4(-03) & 5.0(-01) & 1.4(-08) & & 4.3(-04) & 5.0(-01) & 4.9(-09) & & 1.4(-04) & 5.0(-01) & 1.6(-09) \\ 
1400 & 1.8(-02) & 4.9(-01) & 1.1(-07) & & 5.8(-03) & 5.0(-01) & 5.1(-08) & & 1.8(-03) & 5.0(-01) & 1.9(-08) & & 5.8(-04) & 5.0(-01) & 6.5(-09) \\ 
1500 & 6.2(-02) & 4.7(-01) & 2.0(-07) & & 2.0(-02) & 4.9(-01) & 1.1(-07) & & 6.4(-03) & 5.0(-01) & 5.5(-08) & & 2.0(-03) & 5.0(-01) & 2.1(-08) \\ 
1600 & 1.7(-01) & 4.1(-01) & 4.9(-07) & & 5.9(-02) & 4.7(-01) & 2.2(-07) & & 1.9(-02) & 4.9(-01) & 1.2(-07) & & 6.1(-03) & 5.0(-01) & 5.4(-08) \\ 
1700 & 3.9(-01) & 3.0(-01) & 1.1(-06) & & 1.5(-01) & 4.3(-01) & 4.9(-07) & & 5.0(-02) & 4.8(-01) & 2.2(-07) & & 1.6(-02) & 4.9(-01) & 1.1(-07) \\ 
1800 & 6.8(-01) & 1.6(-01) & 1.9(-06) & & 3.1(-01) & 3.4(-01) & 1.0(-06) & & 1.1(-01) & 4.4(-01) & 4.4(-07) & & 3.7(-02) & 4.8(-01) & 2.0(-07) \\ 
1900 & 8.8(-01) & 5.8(-02) & 2.6(-06) & & 5.5(-01) & 2.2(-01) & 1.8(-06) & & 2.3(-01) & 3.9(-01) & 8.8(-07) & & 7.9(-02) & 4.6(-01) & 3.7(-07) \\ 
2000 & 9.7(-01) & 1.7(-02) & 3.7(-06) & & 7.8(-01) & 1.1(-01) & 2.4(-06) & & 4.0(-01) & 3.0(-01) & 1.5(-06) & & 1.5(-01) & 4.2(-01) & 6.9(-07) \\ 
2100 & 9.9(-01) & 5.1(-03) & 5.3(-06) & & 9.1(-01) & 4.3(-02) & 3.3(-06) & & 6.1(-01) & 1.9(-01) & 2.2(-06) & & 2.7(-01) & 3.7(-01) & 1.2(-06) \\ 
2200 & 1.0(+00) & 1.6(-03) & 6.7(-06) & & 9.7(-01) & 1.5(-02) & 4.6(-06) & & 7.9(-01) & 1.0(-01) & 2.9(-06) & & 4.2(-01) & 2.9(-01) & 1.8(-06) \\ 
2300 & 1.0(+00) & 5.7(-04) & 8.0(-06) & & 9.9(-01) & 5.6(-03) & 6.0(-06) & & 9.1(-01) & 4.7(-02) & 3.9(-06) & & 5.9(-01) & 2.0(-01) & 2.4(-06) \\ 
2400 & 1.0(+00) & 2.2(-04) & 1.1(-05) & & 1.0(+00) & 2.2(-03) & 7.3(-06) & & 9.6(-01) & 2.0(-02) & 5.1(-06) & & 7.5(-01) & 1.2(-01) & 3.2(-06) \\ 
2500 & 1.0(+00) & 9.0(-05) & 1.6(-05) & & 1.0(+00) & 9.0(-04) & 9.0(-06) & & 9.8(-01) & 8.7(-03) & 6.3(-06) & & 8.6(-01) & 6.8(-02) & 4.1(-06) \\ 
2600 & 1.0(+00) & 4.0(-05) & 2.4(-05) & & 1.0(+00) & 4.0(-04) & 1.2(-05) & & 9.9(-01) & 3.9(-03) & 7.7(-06) & & 9.3(-01) & 3.5(-02) & 5.2(-06) \\ 
2700 & 1.0(+00) & 1.9(-05) & 3.4(-05) & & 1.0(+00) & 1.9(-04) & 1.8(-05) & & 1.0(+00) & 1.9(-03) & 9.7(-06) & & 9.7(-01) & 1.7(-02) & 6.3(-06) \\ 
2800 & 1.0(+00) & 9.2(-06) & 4.3(-05) & & 1.0(+00) & 9.2(-05) & 2.6(-05) & & 1.0(+00) & 9.2(-04) & 1.3(-05) & & 9.8(-01) & 8.9(-03) & 7.7(-06) \\ 
2900 & 1.0(+00) & 4.8(-06) & 5.2(-05) & & 1.0(+00) & 4.8(-05) & 3.4(-05) & & 1.0(+00) & 4.8(-04) & 1.8(-05) & & 9.9(-01) & 4.7(-03) & 9.7(-06) \\ 
3000 & 1.0(+00) & 2.6(-06) & 6.3(-05) & & 1.0(+00) & 2.6(-05) & 4.3(-05) & & 1.0(+00) & 2.6(-04) & 2.5(-05) & & 9.9(-01) & 2.6(-03) & 1.3(-05) \\ 
\enddata
\tablecomments{Abundances are calculated by $n$(X)/($n$(H)+2$n$(H$_{2}$)). Values indicated by $a(b)$ mean $a\times10^{b}$. \\
The abundance of hydrogen ion is negligible under the condition of low temperature ($<$4000~K). }
\end{deluxetable}

\begin{deluxetable}{ccccccc}
\tablewidth{0pt}
\tabletypesize{\scriptsize}
\tablecaption{Temperature and total hydrogen density for given turnover frequencies. 
\label{tab-turnfreq}}
\tablehead{
\colhead{} & \multicolumn{2}{c}{Power law fit\tablenotemark{a}} & &
\multicolumn{3}{c}{$n$(H)+2$n$(H$_{2}$) (cm$^{-3}$)} \\
\cline{2-3} \cline{5-7} \\
\colhead{$T$(K)} & \colhead{$a$} & \colhead{$b$} & &
\colhead{200~GHz} & \colhead{300~GHz} & \colhead{600~GHz} 
}
\startdata
1000 & -8.84 & 0.750  & & 7.1(14) & 1.2(15) & 3.1(15) \\
1100 & -8.32 & 0.751  & & 1.4(14) & 2.4(14) & 6.0(14) \\
1200 & -7.91 & 0.753  & & 3.6(13) & 6.2(13) & 1.5(14) \\
1300 & -7.64 & 0.761  & & 1.2(13) & 2.0(13) & 5.0(13) \\
1400 & -7.57 & 0.778  & & 5.0(12) & 8.3(12) & 2.0(13) \\
1500 & -7.61 & 0.798  & & 2.6(12) & 4.3(12) & 1.0(13) \\
1600 & -7.41 & 0.800  & & 1.4(12) & 2.3(12) & 5.5(12) \\
1700 & -7.03 & 0.787  & & 7.2(11) & 1.2(12) & 2.9(12) \\
1800 & -6.70 & 0.777  & & 3.9(11) & 6.6(11) & 1.6(12) \\
1900 & -6.55 & 0.777  & & 2.4(11) & 4.1(11) & 1.0(12) \\
2000 & -6.50 & 0.785  & & 1.7(11) & 2.8(11) & 6.8(11) \\
2100 & -6.55 & 0.797  & & 1.2(11) & 2.1(11) & 5.0(11) \\
2200 & -6.67 & 0.815  & & 1.0(11) & 1.7(11) & 4.0(11) \\
2300 & -6.84 & 0.834  & & 9.0(10) & 1.5(11) & 3.4(11) \\
2400 & -6.95 & 0.849  & & 7.9(10) & 1.3(11) & 2.9(11) \\
2500 & -6.97 & 0.856  & & 6.6(10) & 1.1(11) & 2.4(11) \\
2600 & -6.90 & 0.858  & & 5.4(10) & 8.6(10) & 1.9(11) \\
2700 & -6.80 & 0.856  & & 4.3(10) & 6.9(10) & 1.6(11) \\
2800 & -6.70 & 0.854  & & 3.5(10) & 5.6(10) & 1.3(11) \\
2900 & -6.62 & 0.853  & & 2.9(10) & 4.6(10) & 1.0(11) \\
3000 & -6.59 & 0.856  & & 2.5(10) & 4.0(10) & 8.9(10) \\
\enddata
\tablecomments{Total hydrogen densities, $n$(H)+2$n$(H$_{2}$), are calculated with fixed turnover frequencies and temperatures. 
The path length is fixed to be 84~AU. 
}
\tablenotetext{a}{See equation (\ref{eq-turnfunc}). Values indicated by $a(b)$ mean $a\times10^{b}$. }
\end{deluxetable}

{}

\end{document}